\documentclass[pdflatex,sn-mathphys-num]{sn-jnl}% Math and Physical Sciences Numbered Reference Style 

\usepackage{graphicx}%
\usepackage{multirow}%
\usepackage{longtable}
\usepackage{amsmath,amssymb,amsfonts}%
\usepackage{amsthm}%
\usepackage{mathrsfs}%
\usepackage[title]{appendix}%
\usepackage{xcolor}%
\usepackage{textcomp}%
\usepackage{manyfoot}%
\usepackage{booktabs}%
\usepackage{algorithm}%
\usepackage{algorithmicx}%
\usepackage{algpseudocode}%
\usepackage{listings}%
\usepackage{comment}
\usepackage[colorinlistoftodos]{todonotes}
\theoremstyle{thmstyleone}%
\theoremstyle{thmstyletwo}%

\theoremstyle{thmstylethree}%

\begin{document}

\title[Article Title]{Automatic techniques for issue report classification: A systematic mapping study}

%%=============================================================%%
% GivenName	-> \fnm{Joergen W.}

\author{\fnm{Muhammad} \sur{Laiq} and \fnm{Felix} \sur{Dobslaw }}\email{muhammad.laiq@miun.se, felix.dobslaw@miun.se}

\affil{Mid Sweden University, Department of Communication, Quality Management and Information Systems, Campus Östersund, Sweden}

%Target call: https://link.springer.com/collections/jgagjgcgef 
\abstract{
%\noindent\textbf{Context:} 
Several studies have evaluated automatic techniques for classifying software issue reports to assist practitioners in effectively assigning relevant resources based on the type of issue. Currently, no comprehensive overview of this area has been published. A comprehensive overview will help identify future research directions and provide an extensive collection of potentially relevant existing solutions.
This study aims to provide a comprehensive overview of the use of automatic techniques to classify issue reports.
We conducted a systematic mapping study and identified 46 studies on the topic.

The study results indicate that the existing literature applies various techniques for classifying issue reports, including traditional machine learning and deep learning-based techniques and more advanced large language models. Furthermore, we observe that these studies (a) lack the involvement of practitioners, (b) do not consider other potentially relevant adoption factors beyond prediction accuracy, such as the explainability, scalability, and generalizability of the techniques, and (c) mainly rely on archival data from open-source repositories only. Therefore, future research should focus on real industrial evaluations, consider other potentially relevant adoption factors, and actively involve practitioners.}


\maketitle

\section{Introduction}

In large-scale software development, a large number of issue reports are submitted by end-users, developers, and testers. Issue reports are submitted during development and post-deployment as part of the maintenance and further development of services/products. In the next step, these reports are assigned to the relevant developers. Developers then use the submitted information to recreate the issue, identify root causes, and fix it. This process can be time-consuming and resource-intensive and can involve back-and-forth discussions among several stakeholders. Moreover, these reports vary significantly in nature. Some reports may be submitted due to misunderstandings, while others might suggest improvements to the documentation or request new features. Some reports describe issues that require a code fix \cite{antoniol2008bug}. Consequently, each issue report requires a different action. Thus, practitioners need the ability to identify the type of a newly submitted issue report early in the issue management process to effectively assign the relevant resources.

Manually identifying and classifying issue reports is labor intensive and error-prone \cite{herzig2013s}. Thus, several automatic techniques (e.g., \cite{zanetti2013categorizing,laiq2022early,cho2022classifying,siddiq2022bert}) have been evaluated to assist practitioners in automatically classifying issue reports into relevant categories such as bugs (requiring a code correction) and non-bugs (all other categories, e.g., feature request and question).

Currently, no comprehensive overview of this area has been published. A comprehensive overview will provide an extensive collection of potentially relevant existing solutions and help identify future research directions.

In response, we  conducted a systematic mapping study \cite{petersen2008systematic} to identify the primary studies on the topic, answering the following overarching research question:
\textit{\textbf{RQ: Which techniques have been proposed in the literature for classifying issue reports?}}
Following a systematic approach, we developed a comprehensive map that covers various aspects, including the study's proposed technique, evaluation approach, evaluation metrics, and the used issue report features.

The main contributions of this mapping study are as follows.
\begin{itemize}

    \item A comprehensive overview/map of the existing work on issue report classification with detailed information about the studies.
    
    \item Several insights from the existing literature on issue report classification. For instance, BERT-based models, such as RoBERTa, are recognized as state-of-the-art for this task. Furthermore, traditional ML techniques such as logistic regression and SVM can achieve prediction performance comparable to complex deep learning methods and BERT-based models in classifying issue reports.
    
    \item Several potential future research directions for issue report classification. For example, the need for evaluations of the proposed techniques in real industrial settings and a careful assessment of the reliability of ground-truth data used for training and evaluating ML models in issue report classification.
\end{itemize}

This paper is structured as follows. Section \ref{sec:back-related-work} describes the background knowledge and related work on the topic. Section \ref{sec:research-method} describes the search strategy, selection criteria, data extraction, data analysis and classification, and validity threats to this study. The results are presented in Section~\ref{sec:results}. Section~\ref{sec:discussion} discusses the overall findings, while Section~\ref{sec:conclusion} concludes the paper.
 
\section{Background and related work}\label{sec:back-related-work}

Issue management is a costly and complicated process that includes several activities such as reporting, assigning, and resolving issues \cite{zou2018practitioners,parnin2011automated,laiq2023data}. 
In large projects, developers, testers, and end-users submit a large number of issue reports \cite{fan2018chaff,zhang2015survey}. An issue report\footnote{An issue report is also referred to as defect report, trouble report, bug report, failure report, error report, or incident report.} is an essential software artifact and plays an important role in the debugging and fixing of issues \cite{zou2018practitioners}. To assist practitioners in managing the workload of issue reports and streamlining the issue-fixing processes, significant research in software engineering has focused on automated techniques \cite{zou2018practitioners} such as issue assignee recommendation \cite{borg2024adopting}, bug localization \cite{kim2013should}, and duplicate issue report detection \cite{zhang2023duplicate}.

Similarly, to assist practitioners in assigning resources effectively to issue reports, several studies have investigated the use of automatic techniques to categorize issue reports, such as into \textit{bugs} (those that require code correction) and \textit{non-bugs} (e.g., feature requests, questions, and documentation) \cite{antoniol2008bug,herzig2013s,zanetti2013categorizing,colavito2024large}. Several automatic techniques have been evaluated in the literature for this task, including traditional ML \cite{fan2018chaff,laiq2022early} and deep learning-based techniques \cite{he2020deep, kwak2023multimodal} and more advanced large language models \cite{siddiq2022bert,heo2024comparison,laiq2024industrial}.

While a number of primary studies have been conducted on issue report classification, to the best of our knowledge, no comprehensive overview has been published. Such an overview would help identify future research directions and provide an extensive collection of relevant existing solutions. K{\"o}ksal and {\"O}zt{\"u}rk \cite{koksal2022survey} reviewed the literature on issue report classification and identified 18 primary studies. However, only 7 of these studies focus on classifying issue reports into bugs and non-bugs. Furthermore, their review is not systematic in nature \cite{kitchenham2016evidence}. In addition, it only reports limited information about primary research, i.e., the techniques used, the number of issue reports, and their sources (i.e., proprietary or open source data). Our review provides (a) a comprehensive/map overview of the area following a systematic approach \cite{petersen2008systematic}, (b) detailed information on primary research, (c) key lessons learned from the existing work, and (d) future research directions.

Moreover, several reviews related to issue management have been conducted in the literature. For example, Li et al. \cite{zhang2015survey} reviewed the literature on issue report analysis. Zhou et al. \cite{zou2018practitioners} reviewed the literature on automatic techniques to manage issue reports and conducted a survey with practitioners to gain their insights. Gomes et al. \cite{gomes2019bug} reviewed the literature on predicting the severity of issue reports. Uddin et al. \cite{uddin2017survey} reviewed the literature on prioritization of issue reports. 
While these studies relate to our work, they do not answer the research question posed in this study. Therefore, we conducted a systematic mapping study \cite{petersen2008systematic} to answer the posed research question.

\section{Research method} \label{sec:research-method}
\begin{figure}[!ht]
    \centering
    \includegraphics[width=\columnwidth]{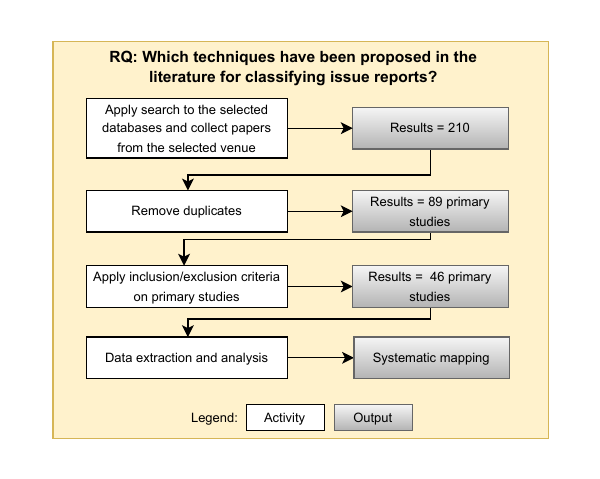}
    \vspace{-4mm}
    \caption{Overview of our approach}
    \label{fig:sms-process}
\end{figure}

We conducted a systematic mapping study \cite{petersen2008systematic} to answer the research questions listed in Table \ref{tab:research-questions}. Figure~\ref{fig:sms-process} provides an overview of our research approach.

\begin{table*}[ht]
    \centering
    \small
    \caption{Research questions}
    \label{tab:research-questions}
    \begin{tabular}{p{6cm}p{6cm}}
    \toprule
     \textbf{Research question} & \textbf{Sub-research questions} 
    \\ \midrule 
   \multirow{7}{6cm}{\textbf{RQ: Which techniques have been proposed in the literature for classifying issue reports?} With this research question, we aim to identify proposed automatic techniques for issue report classification. In the following research questions, we further collect information about the pre-processing techniques, the evaluation approaches and metrics, and other related information, such as the context of the study.}
    & \textbf{RQ1:} Which automatic techniques have been proposed? \\ 
    & \textbf{RQ2:} Which features have been used? \\
    & \textbf{RQ3:} Which pre-processing techniques have been used? \\
    & \textbf{RQ4:} Which evaluation approaches have been used? \\
    & \textbf{RQ5:} Which evaluation metrics have been used? \\ 
    & \textbf{RQ6:} In which contexts has research been conducted? \\
    \bottomrule
    \end{tabular}

 \end{table*}  

\subsection{Search strategy}\label{sec:search-strategy}

Table \ref{tab:searchstring} shows our search string in selected databases to find relevant literature on automatically classifying issue reports.
Our search string focuses on studies that classify issue reports according to the classification criteria as defined by Antoniol et al. \cite{antoniol2008bug} and Herzig et al. \cite{herzig2013s}:  (a) those that require code fixes, that is, \textit{bugs}, and (b) \textit{non-bugs}, e.g., documentation, feature requests, and questions.

In addition to searching relevant papers in the selected databases, we complemented our search strategy by manually collecting articles from a venue\footnote{First addition: https://nlbse2022.github.io/program/} that focuses on issue report classification: the Workshop on NL-based Software Engineering co-located with the International Conference on Software Engineering. We collected papers from all the additions of the workshop from 2022 to 2024.

Table \ref{tab:db-hits} shows the number of search results in each database and the number of collected papers from the selected venue. The search was performed on December 7, 2024.

\begin{table}[ht]
    \centering
    \caption{Search strings in selected databases}
    \vspace{-2mm}
    \label{tab:searchstring}
    \begin{tabular}{p{2.3cm}p{9.7cm}}
    \toprule
     \textbf{Database/Venue} & \textbf{Search string}
    \\ \midrule 
     Scopus &  \textit{TITLE-ABS-KEY(((``bug report classification" OR ``defect report classification" OR ``issue report classification") OR (``classifying bug report" OR ``classifying defect report" OR ``classifying issue report") OR ((predict OR determine OR identify) AND (``valid bug reports" OR ``invalid bug reports"))))} 
     \\ \midrule
    
    Web of Science & Search was performed on title, abstract, and keywords and combined using the OR operator: \textit{((``bug report classification" OR ``defect report classification" OR ``issue report classification") OR (``classifying bug report" OR ``classifying defect report" OR ``classifying issue report") OR ((predict OR determine OR identify) AND (``valid bug reports" OR ``invalid bug reports")))} 
    \\ \midrule
    
    IEEE Xplore &  Full-text search was performed using the command feature of IEEE Xplore: \textit{((``bug report classification" OR ``defect report classification" OR ``issue report classification") OR (``classifying bug report" OR ``classifying defect report" OR ``classifying issue report") OR ((predict OR determine OR identify) AND (``valid bug reports" OR ``invalid bug reports")))}
    \\ \midrule
    
    ACM  & Search was performed on title, abstract, and keywords and manually combined:\textit{((``bug report classification" OR ``defect report classification" OR ``issue report classification") OR (``classifying bug report" OR ``classifying defect report" OR ``classifying issue report") OR ((predict OR determine OR identify) AND (``valid bug reports" OR ``invalid bug reports")))}
    \\ \midrule
    
    Selected venue & Papers were collected from the selected venue on issue report classification from 2022 -- 2024, i.e., Workshop on NL-based Software Engineering, co-located with the International Conference on Software Engineering.
    \\ \bottomrule

    \end{tabular}

\end{table}

\begin{table}[ht]
    \centering
    \caption{Search results in the selected databases and venue}
    \vspace{-2mm}
    \label{tab:db-hits}
    \begin{tabular}{p{4cm}p{3cm}}
    \toprule
     \textbf{Database} & \textbf{Search results} 
    \\  \midrule 
    Scopus & 97
    \\  \midrule
    
    Web of Science & 50
    \\ \midrule
    
    IEEE Xplore & 39
    \\  \midrule
    
    ACM & 13
    \\ \midrule    
    
    Selected venue & 11
    \\ \midrule
    
     & \textbf{Total = 210}
    \\ \bottomrule

    \end{tabular}
\end{table}

\subsection{Study selection and outcome}\label{sec:selection-criteria}

We selected papers based on title, abstract reading, and full-text reading. We read full-text only when we could not decide whether to include or exclude a paper based on the title and abstract reading. The following inclusion criteria were followed for the study selection:

\begin{itemize}

\item [] \textbf{I1:} The paper is peer-reviewed.

\item [] \textbf{I2:} The paper is written in English.

\item [] \textbf{I3:} The paper focuses on classifying issue reports according to the classification criteria as defined by Antoniol et al. \cite{antoniol2008bug} and Herzig et al. \cite{herzig2013s}:  (a) those that require code fixes, that is bugs. (b) non-bugs, e.g., documentation, feature requests, questions. 

\end{itemize}

Figure \ref{fig:sms-process} shows the number of included and excluded papers at each stage. The first author applied the selection criteria mentioned above to all 89 papers. As a result, 46 papers were selected for data extraction. See an overview in Figure \ref{fig:sms-process}.

In total, 43 papers were excluded. The same reviewer, the first author, reconsidered all 43 excluded papers after a gap of 7 weeks from the first iteration of selecting papers to reduce the risk of excluding a relevant paper.
In addition, the list of the excluded papers at this stage is made available at \url{https://tinyurl.com/j92hvesa} for transparency.

\subsection{Data extraction}\label{sec:data-extraction}

Table \ref{tab:dex-template-1} shows the data extraction template we used to extract data from the selected primary studies.
The first author performed data extraction on all 46 primary studies. The second author checked the data extraction for five randomly selected primary studies. No discrepancies were found between the two authors.

\begin{table}[!ht]
    \centering
    \caption{Data extraction template}
    \vspace{-4mm}
    \label{tab:dex-template-1}
    \begin{tabular}{lp{8.5cm}c}
    \toprule
     \textbf{Item} & \textbf{Description} & \textbf{Mapping to RQs}
    \\ \midrule 
     
    \#1 & Meta information, e.g., the title of the paper, publication year, and type of paper, i.e., conference/journal/workshop. & --
    \\ \midrule
    
    \#2 & Proposed automatic technique, e.g., Logistic regression.  & RQ1
    \\ \midrule
    
    \#3 & Used features with a technique, e.g., title and description of an issue report.  & RQ2
    \\ \midrule
    
    \#4 & Used pre-processing techniques, e.g., Word2vec and Term frequency-inverse document frequency (TF-IDF).  & RQ3
    \\ \midrule
    
    \#6 & Used evaluation approach, e.g., 10-fold cross-validation and test on a fixed test set.  & RQ4
    \\ \midrule
    
    \#5 & Used evaluation metrics, e.g., F1-score and accuracy.  & RQ5
    \\ \midrule
    
    \#7 & Study context: (a) study realism: open-source (OSS), closed-source (CSS), and (b) the source of data, e.g., Github and Bugzilla.  & RQ6
    \\  \bottomrule

    \end{tabular}

\end{table}

\subsection{Analysis and classification}\label{sec:analysis-and-clasification}

The extracted information was tabulated and visualized using graphs. We also categorized the proposed techniques into higher-level categories, such as traditional ML and deep learning techniques and BERT-based large language models. Similarly, we grouped studies based on evaluation approaches, evaluation metrics, and pre-processing techniques.

\subsection{Validity threats}\label{sec:validity-threats}

We used the guidelines by Ampatzogloua et al. \cite{ampatzoglou2019identifying} to consider the potential validity threats to our study and identify mitigation actions. In the following subsections, we discuss some of the validity threats and our mitigation actions.

\subsubsection{Study selection validity}\label{sec:study-selection-validity}

A potential threat to this mapping study is the missing of relevant studies on issue report classification. We mitigated this by carefully developing our search string that focuses on identifying relevant studies on issue report classification. In addition, we have used two indexing (i.e., Scopus and Web of Science) and two publisher-specific databases (i.e., IEEE Xplore and ACM) without restricting our search to papers from specific years to have sufficient coverage. These databases sufficiently cover the software engineering literature \cite{kitchenham2016evidence,usman2023quality}.
Moreover, we also complemented our search strategy by manually identifying relevant studies from a relevant venue dedicated to the topic (see details in Section \ref{sec:search-strategy}).

Another threat to this review is bias during the paper selection process. We mitigated this risk by ensuring that every excluded paper was reconsidered after a 7-week gap by the same reviewer. A similar approach (i.e., test-retest) has been applied and recommended when a single reviewer conducts steps of systematic reviews or mapping studies \cite{kitchenham2010s, petersen2011measuring}.

\subsubsection{Data validity}\label{sec:data-validity}

Another potential threat to this study is bias in the data extraction and analysis of primary studies. Two reviewers were involved in the data extraction and analysis process to mitigate this threat. The first author performed data extraction on all 46 studies. To increase confidence in data extraction, the second author reviewed a randomly selected subset of five primary studies to verify the correctness of the extracted data. No discrepancies were found between the two authors.

A limitation of this study is that we did not conduct a quality assessment of the included primary studies. This is not a major threat to the results of this mapping, where we show what techniques have been used for issue report classification. However, any synthesis of evidence to ascertain what works, how, when, and for whom \cite{Ali16} would require consideration of the context and the strength of the evidence.

\subsubsection{Research validity}\label{sec:research-validity}

The research method chosen for this study is a systematic mapping study based on our aim, i.e., to structure and provide an overview of existing work on issue report classification. Before conducting this review, we developed a review protocol following the systematic mapping guidelines \cite{petersen2008systematic,usman2023quality}. More than one researcher was involved in the design and execution of the study, and the protocol was iteratively improved before the final execution of the review process.

The data collected in the process of this study has been made available for transparency and replication (including the list of primary studies, the list of papers excluded from the review after full-text reading, and a complete map in the form of a spreadsheet that readers can use to navigate the literature map).

\section{Results and analysis} \label{sec:results}

This study provides a comprehensive map of various automatic techniques for issue report classification. The complete map, too large to reproduce here, including metadata and other contextual information, can be found at \url{https://tinyurl.com/wdkdxj65}; a simplified version can be found in Appendix \ref{appendix:complete-map}. Table \ref{tab:excerpt-map} shows an excerpt of the simplified mapping of the automatic techniques for the classification of issue reports developed in this study.

In the following subsections, we first describe the characteristics of the selected primary studies and then present the results of our research questions RQ1--RQ6 stated in Table \ref{tab:research-questions}.

\begin{figure}[h]
    \centering
    \includegraphics[width=\columnwidth]{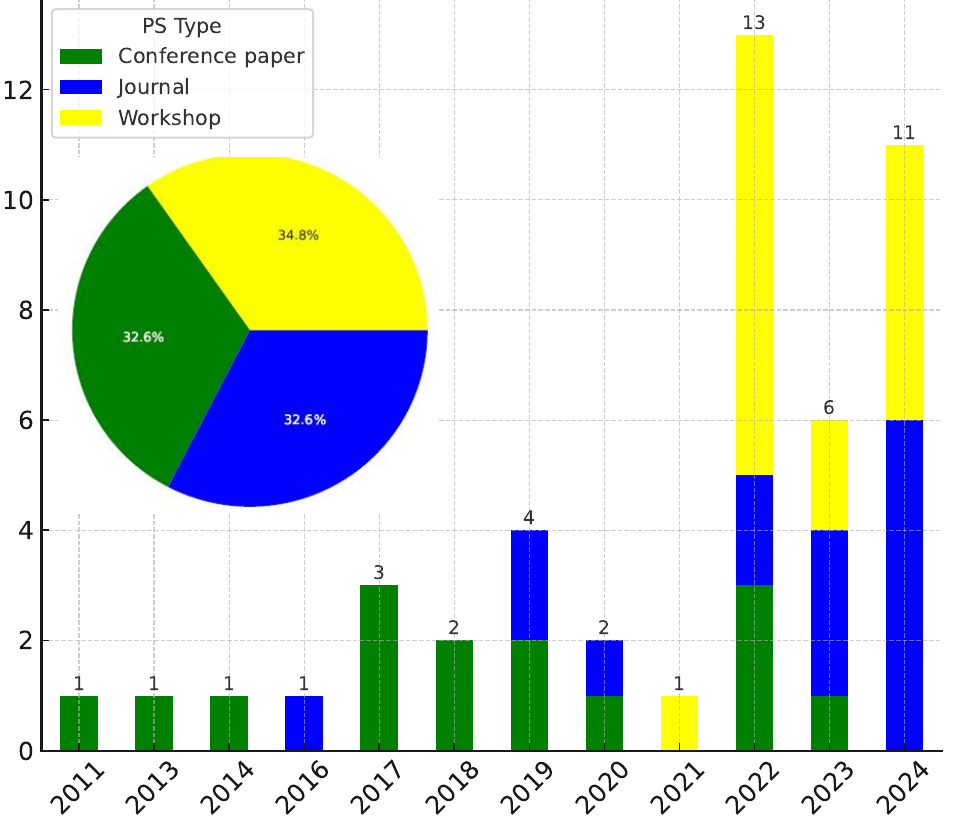}
    \vspace{-4mm}
    \caption{Yearly distribution of primary studies grouped by their publication venue type}
    \label{fig:yearly-distirbution}
\end{figure}

\noindent\textbf{Characteristics of the analyzed primary studies:} Figure \ref{fig:yearly-distirbution} shows the yearly distribution between 2011 and 2024 of primary studies on issue report classification. These studies are published in leading software engineering venues, including the Empirical Software Engineering Journal, the International Conference on Software Engineering, and the Journal of Systems and Software.

As shown in Figure \ref{fig:yearly-distirbution}, 16 (34.8\%) of the studies belonged to workshops, 15 (32.6\%) of the studies belonged to conferences, and another 15 (32.4\%) of the studies belonged to journals.

\subsection*{RQ1: Which automatic techniques have been proposed?}

\begin{figure}[!ht]
    \centering
    \includegraphics[width=\columnwidth,height=0.5\paperheight]{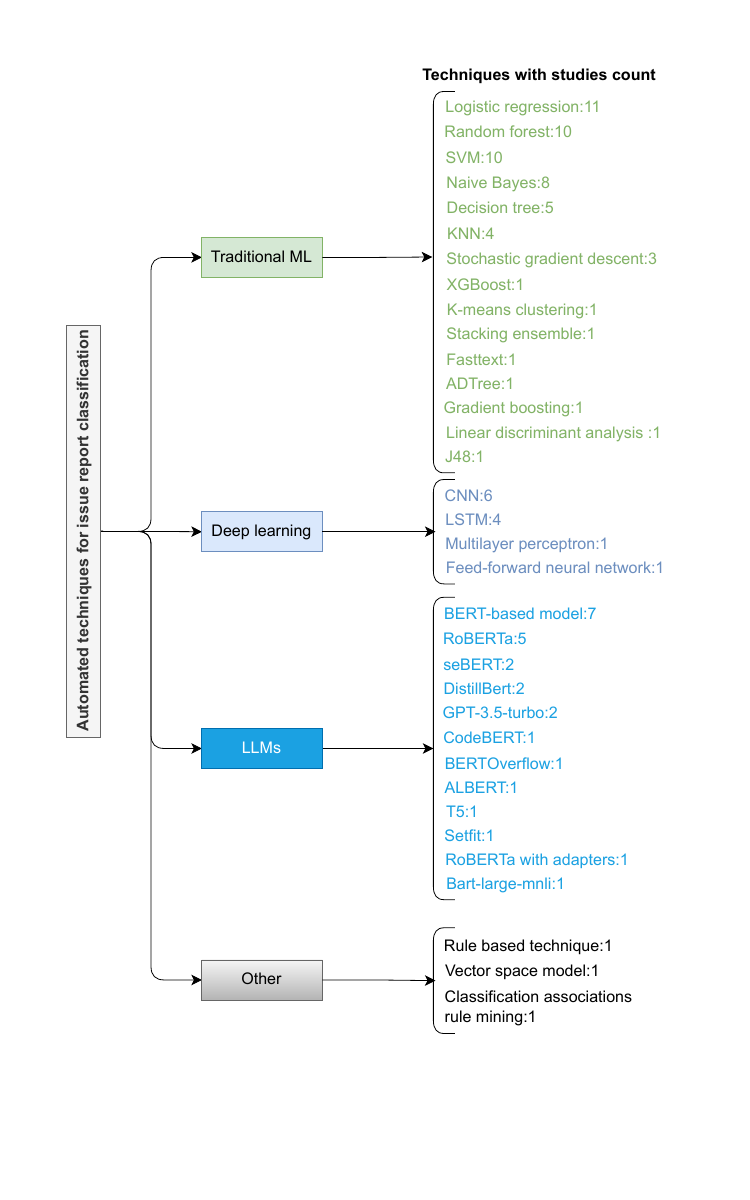}
    \caption{Categorization of automatic techniques for issue report classification}
    \label{fig:autoamtic-technique}
\end{figure}

Table \ref{tab:techniques} and Figure \ref{fig:autoamtic-technique} provide an overview of the proposed automatic techniques in the literature for issue report classification.

Traditional ML techniques are predominant, with logistic regression at the top used in 11 studies, followed by random forest and SVM used in 10 studies each. Naive Bayes has been used in 8 studies, while decision trees have been used in 5 studies. Stochastic gradient boosting and XGBoost have been used in 3 and 2 studies, respectively. All other traditional ML techniques have been used in a single study.

Deep learning techniques have also been used in several studies, with convolutional neural networks used in 6 studies and long short-term memory networks in 4 studies. 

Transformer-based models, particularly BERT and its variants, are becoming increasingly popular, used in 7 studies, followed by RoBERTa, which has been used in 5 studies, along with other specialized adaptations like seBERT and DistillBERT.

Few-shot and zero-shot learning approaches are emerging trends, with five studies evaluating them for issue report classification. The text-to-text transfer transformer (T5) model has been evaluated in a single study.

\begin{scriptsize}
\begin{longtable}{p{1.9cm}p{3cm}p{4.7cm}p{1.8cm}}
    \caption{Automatic techniques proposed in the literature for classifying issue reports}
    \label{tab:techniques}
\\ \toprule
\textbf{Category} & \textbf{Technique} & \textbf{Relevant studies} & \textbf{Count}
 \\ \toprule
\endfirsthead
\toprule 

\textbf{Category (continued)} & \textbf{Technique (continued)} & \textbf{Relevant studies \newline(continued)} & \textbf{Count \newline(continued)}
\\
\toprule
\endhead
\toprule
\endfoot
\toprule
\endlastfoot

\multirow{1}{4cm}{Traditional \newline ML}
& Logistic regression & PS6, PS10, PS13, PS20, PS24, PS26, PS30, PS32, PS41, PS42, PS44 & 11             \\ \cmidrule{2-4}

& Random forest & PS2, PS6, PS9, PS10, PS13, PS16, PS32, PS41, PS42, PS44 & 10 
\\ \cmidrule{2-4}

& Support vector machine (SVM) & PS2, PS6, PS9, PS10, PS15, PS16, PS26, PS32, PS32, PS42  & 10
 \\ \cmidrule{2-4}

& Naive Bayes & PS6, PS9, PS10, PS20, PS24, PS30, PS41, PS42, PS44 & 9 
\\ \cmidrule{2-4}

& Decision tree  & PS10, PS20, PS26, PS30, PS41  & 5 
\\ \cmidrule{2-4}

& K-nearest neighbors (KNN)  & PS5, PS10, PS26, PS38 & 4
\\ \cmidrule{2-4}

& Stochastic gradient descent (SGD) & PS6, PS7, PS41 & 3
\\ \cmidrule{2-4}

& XGBoost & PS32, PS44 & 2
\\ \cmidrule{2-4}

& K-means clustering  & PS2 & 1
\\ \cmidrule{2-4}

& Stacking ensemble  & PS6 & 1
\\ \cmidrule{2-4}

& Fasttext & PS22 & 1
\\ \cmidrule{2-4}

& ADTree & PS24 & 1
\\ \cmidrule{2-4}

& Gradient boosting & PS41 & 1
\\ \cmidrule{2-4}

& Linear discriminant analysis  & PS41 & 1
\\ \cmidrule{2-4}

& J48 & PS44 & 1             
\\ \midrule

\multirow{4}{4cm}{Deep \newline learning}
& Convolutional neural network (CNN) & PS4, PS21, PS25, PS32, PS33, PS37 & 6 
\\ \cmidrule{2-4}

& Long short-term memory network (LSTM)  & PS19, PS21, PS37, PS39  & 4
\\ \cmidrule{2-4}
 
& Multilayer perceptron  & PS6 & 1
\\ \cmidrule{2-4}

& Feed-forward neural network & PS36  & 1
\\ \midrule

\multirow{7}{2cm}{BERT-based models} 
& BERT-based model & PS3, PS12, PS18, PS31, PS32, PS34, PS43 & 7
\\ \cmidrule{2-4}

& RoBERTa & PS3, PS11, PS17, PS34, PS46 & 5
\\ \cmidrule{2-4}

& seBERT & PS3, PS45 & 2
\\ \cmidrule{2-4}

& DistillBert & PS27, PS32 & 2
\\ \cmidrule{2-4}

& CodeBERT & PS3 & 1
\\ \cmidrule{2-4}

& BERTOverflow & PS3 & 1
\\ \cmidrule{2-4}

& ALBERT & PS34 & 1
\\ \midrule

T5 & Text-to-text transfer transformer  & PS40  & 1 
\\ \midrule

%GPT & GPT-3.5-turbo & PS8 & 1
%\\ \midrule

\multirow{2}{4cm}{Few shot \newline learning}
& Setfit & PS29, PS35 & 2
\\ \cmidrule{2-4}

& GPT-3.5-turbo & PS8 & 1
\\ \cmidrule{2-4}

& RoBERTa with adapters & PS28  & 1
\\ \midrule

\multirow{2}{4cm}{Zero shot \newline learning}
& Bart-large-mnli & PS27 & 1 
\\ \cmidrule{2-4}

& GPT3.5-turbo & PS35 & 1 
\\ \midrule

\multirow{3}{*}{Other} & Rule based technique & PS1 & 1
\\ \cmidrule{2-4}

& Vector space model & PS14 & 1 \\ \cmidrule{2-4}

& Classification associations rule mining & PS23  &  1
\\ \bottomrule

\end{longtable}
\end{scriptsize}

\subsection*{RQ2: Which features have been used?}
Table \ref{tab:features-set} shows a summary of the features used in the literature for issue report classification. As shown in the table, the description/body and title/heading of the issue reports, are the most used features in the literature, used in 38 and 32 studies, respectively. Other notable features include priority (6 studies), severity (6 studies), summary (4 studies), and component (4 studies), each appearing in 4 to 6 studies. Features related to submitter experience, collaboration networks, and comments on issue reports were utilized in 3 studies. All other features were used in 1 to 2 studies.

\begin{scriptsize}
\begin{longtable}{p{3.8cm}p{6.8cm}p{1.8cm}}
    \caption{Features used with techniques for issue report classification}
    \label{tab:features-set}
\\ \toprule
\textbf{Feature} & \textbf{Relevant studies} & \textbf{Count}
 \\ \toprule
\endfirsthead
\toprule 

\textbf{Feature (continued)} & \textbf{Relevant studies (continued)} & \textbf{Count \newline(continued)}
\\
\toprule
\endhead
\toprule
\endfoot
\toprule
\endlastfoot

        Description/Body & PS1, PS3, PS6, PS7, PS8, PS9, PS10, PS11, PS12, PS13, PS14, PS16, PS17, PS18, PS19, PS20, PS21, PS22, PS23, PS25, PS26, PS27, PS28, PS29, PS30, PS31, PS32, PS34, PS35, PS36, PS37, PS39, PS40, PS41, PS42, PS43, PS45, PS46 & 38 \\ \midrule
        
        Title/Heading & PS1, PS3, PS6, PS7, PS8, PS9, PS11, PS12, PS13, PS16, PS17, PS18, PS19, PS20, PS21, PS22, PS23, PS26, PS28, PS29, PS31, PS32, PS34, PS35, PS36, PS37, PS39, PS40, PS42, PS43, PS45, PS46 & 32 \\ \midrule
        
        Priority & PS5, PS24, PS26, PS32, PS38, PS44 & 6 \\ \midrule
        
        Summary & PS2, PS5, PS24, PS25, PS38, PS41 & 6 \\ \midrule
        
        Severity & PS5, PS24, PS38, PS44 & 4 \\ \midrule
        
        Component & PS5, PS24, PS38, PS44 & 4 \\ \midrule
        
        Submitter & PS5, PS18, PS24 & 3 \\ \midrule
        
        Comments & PS16, PS20, PS36 & 3 \\ \midrule
        Submitter experience, e.g., total submitted issue reports & PS16, PS26, PS32 & 3 \\ \midrule
        
        Reproducibility & PS5, PS24 & 2 \\ \midrule
        
        Features based on collaboration graphs of issue submitters, e.g., the clusters of highly connected people & PS15, PS16 & 2 \\ \midrule
        
        Text completeness & PS16, PS32 & 2 \\ \midrule
        
        Text readability & PS16, PS44 & 2 \\ \midrule
        
        Assignee & PS5, PS24 & 2 \\ \midrule
        
        Operating system & PS5, PS24 & 2 \\ \midrule
        
        Code & PS4, PS33 & 2 \\ \midrule
        
        Image & PS4, PS33 & 2 \\ \midrule
        
        Metadata-based features, e.g., app rating and category & PS27 & 1 \\ \midrule
        
        Issue-author association & PS34 & 1 \\ \midrule
        
        Text sentiment & PS44 & 1 \\ \midrule
        
        User manual & PS21 & 1 \\ \bottomrule

\end{longtable}
\end{scriptsize}

\subsection*{RQ3: Which pre-processing techniques have been used?}

Table \ref{tab:Pre-processing-techniques} presents the reported pre-processing techniques employed for issue report classification across various studies. The table indicates that TF-IDF is the most commonly used approach, appearing in 13 studies. Modern transformer-based methods, like BERT-based and RoBERTa-based tokenizers, have been used in 7 and 4 studies, respectively. The Skip-gram model has also been used in 7 studies. The sentence-transformers model (All-mpnet-base-v2) was used in 2 studies. Two studies also selected textual features manually. All other pre-processing techniques were used in a single study.
\begin{table}[!ht]
    \centering
    \caption{Pre-processing techniques used for issue report classification}
    \begin{tabular}{p{5cm}p{5.8cm}p{0.8cm}}
    \toprule
        \textbf{Pre-processing technique} & \textbf{Relevant studies} & \textbf{Count} \\ \midrule
        
        TF-IDF & PS1, PS2, PS5, PS6, PS7, PS10, PS13, PS14, PS16, PS26, PS30, PS38, PS42 & 13 \\ \midrule
        
        BERT-based tokenizer & PS3, PS12, PS18, PS27, PS31, PS43, PS45 & 7 \\ \midrule
        Skip-gram model & PS19, PS21, PS22, PS25, PS32, PS39, PS41 & 7 \\ \midrule
        RoBERTa-based tokenizer & PS11, PS17, PS28, PS46 & 4 \\ \midrule
        Embedding layer & PS4, PS21, PS33 & 3 \\ \midrule
        Manuall selection of textual features & PS9, PS23 & 2 \\ \midrule
        All-mpnet-base-v2 & PS29, PS35 & 2 \\ \midrule
        AutoTokenizer from transformers & PS34 & 1 \\ \midrule
        Text-embedding-ada-002 & PS36 & 1 \\ \midrule
        T5-based tokenizer & PS40 & 1 \\ \midrule
        GPT-based tokenizer & PS8 & 1 \\ \midrule
        LDA & PS20 & 1 \\ \midrule
        GloVe & PS21 & 1 \\ \bottomrule
    \end{tabular}
    \label{tab:Pre-processing-techniques}
\end{table}

\subsection*{RQ4: Which evaluation approaches have been used?}

Table \ref{tab:evaluation-approaches} presents the evaluation approaches used in the reviewed studies for classifying issue reports. The most common strategy observed is testing on a fixed set, utilized in 27 studies. Cross-validation techniques also feature prominently, with 10-fold cross-validation appearing in 8 studies and 5-fold cross-validation in 5 studies. Additionally, hybrid approaches, such as time split (sorting issue reports data chronologically before splitting data into train and test sets) with 10-fold cross-validation and time split with testing on a fixed set, were used in 3 and 2 studies, respectively. Stratified 10-fold cross-validation was used in a single study.

\begin{table}[!ht]
    \centering
    \caption{Evaluation approaches used for issue report classification}
    \begin{tabular}{p{4.8cm}p{6cm}p{0.9cm}}
    \toprule
        \textbf{Evaluation strategy} & \textbf{Relevant studies} & \textbf{Count} \\ \midrule
        Test on a fixed set & PS1, PS2, PS3, PS4, PS5, PS7, PS8, PS11, PS12, PS13, PS15, PS17, PS18, PS23, PS28, PS29, PS30, PS31, PS33, PS33, PS35, PS38, PS40, PS41, PS43, PS45, PS46 & 27 \\ \midrule
        10-fold CV & PS13, PS20, PS22, PS24, PS37, PS39, PS42, PS44 & 8 \\ \midrule
        5-fold CV & PS9, PS21, PS27, PS10, PS36 & 5 \\ \midrule
        Time split with 10-fold CV & PS16, PS26, PS32 & 3 \\ \midrule
        Time split with test on a fixed set  & PS19, PS25 & 2 \\ \midrule
        Stratified 10-fold CV & PS6 & 1 \\ \bottomrule
    \end{tabular}
    \label{tab:evaluation-approaches}
\end{table}

\subsection*{RQ5: Which evaluation metrics have been used?}

Table \ref{tab:evaluation-metrics} summarizes the evaluation metrics used in the literature for classifying issue reports. F1-score is the most widely used metric overall, appearing in 40 studies, followed by precision and recall, each used in 35 studies. Micro F1-score has been used in 13 studies, and micro recall and micro precision in 11 studies each. Accuracy has been used in 8 studies. AUC and MCC are less common, reported in 6 and 5 studies, respectively. G-measure is the least used, appearing in only 2 studies.
\begin{table}[!h]
    \centering
    \caption{Evaluation metrics used for issue report classification}
    \begin{tabular}{p{3cm}p{7.7cm}p{0.8cm}}
    \toprule
        \textbf{Evaluation metrics} & \textbf{Relevant studies} & \textbf{Count} \\ \midrule
        F1-score & PS2, PS4, PS5, PS6, PS7, PS8, PS10, PS11, PS12, PS13, PS15, PS16, PS17, PS18, PS19, PS20, PS21, PS22, PS24, PS25, PS26, PS27, PS28, PS29, PS30, PS31, PS32, PS33, PS34, PS35, PS36, PS37, PS39, PS40, PS41, PS42, PS43, PS44, PS45, PS46 & 40 \\ \midrule
        Precision & PS1, PS2, PS4, PS5, PS7, PS8, PS11, PS12, PS14, PS15, PS16, PS17, PS18, PS19, PS21, PS22, PS24, PS25, PS26, PS27, PS28, PS29, PS30, PS31, PS32, PS33, PS34, PS35, PS36, PS40, PS42, PS43, PS44, PS45, PS46 & 35 \\ \midrule
        Recall & PS1, PS2, PS4, PS5, PS7, PS8, PS11, PS12, PS14, PS15, PS16, PS17, PS18, PS19, PS21, PS22, PS24, PS25, PS26, PS27, PS28, PS29, PS30, PS31, PS32, PS33, PS34, PS35, PS36, PS40, PS42, PS43, PS44, PS45, PS46 & 35 \\ \midrule
        Micro average F1-score & PS3, PS7, PS8, PS12, PS17, PS28, PS29, PS31, PS34, PS35, PS40, PS45, PS46 & 13 \\ \midrule
        Micro average recall & PS7, PS8, PS12, PS17, PS28, PS29, PS31,PS34, PS40, PS45, PS46 & 11 \\ \midrule
        Micro average precision & PS7, PS8, PS12, PS17, PS28, PS29, PS31, PS34, PS40, PS45, PS46 & 11 \\ \midrule
        Accuracy & PS6, PS9, PS14, PS23, PS32, PS36, PS38, PS41 & 8 \\ \midrule
        AUC & PS16, PS25, PS26, PS32, PS37, PS39 & 6 \\ \midrule
        MCC & PS6,PS26, PS32, PS37, PS39 & 5 \\ \midrule
        G-measure & PS37, PS39 & 2 \\ \bottomrule
    \end{tabular}
    \label{tab:evaluation-metrics}
\end{table}

\subsection*{RQ6: In which contexts has research been conducted?}

Tables \ref{tab:context-of-studies} and \ref{tab:data-sources} provide an overview of the context in which studies on issue classification were conducted. It highlights that most studies, 42, were based on OSS projects, while only a few, 3, were conducted in CSS contexts. The involvement of practitioners in these studies was minimal, with 42 studies reporting no participation from practitioners, and only four studies included them. 

Additionally, 41 studies did not consider important adoption factors for ML techniques beyond prediction accuracy, such as explainability, generalizability, maintainability, and scalability \cite{rana2014framework,paleyes2022challenges}. Only five studies considered these adoption factors for ML techniques, and among those, three focused solely on explainability, which was approached in a limited manner and lacked evaluation. Only two studies thoroughly investigated the adoption factors for ML techniques \cite{rana2014framework,paleyes2022challenges}.

\begin{table}[h!]
\centering
\caption{Context of studies}
\begin{tabular}{p{2.2cm}p{3.5cm}p{5.8cm}}
\toprule
\textbf{Study realism} & \textbf{Practitioners involved} & \textbf{Adoption factors considered} 
\\ \midrule

OSS: 42  & No: 42 & No: 41 
\\ \midrule

CSS: 3   & Yes: 4  & Yes: 2 
\\ \midrule

& & Only explainability, limited, no evaluation: 3 
\\ \bottomrule
\end{tabular}
\label{tab:context-of-studies}
\end{table}

\begin{table}[h!]
\centering
\caption{Data Sources and relevant Studies}
\begin{tabular}{lp{7.2cm}c}
\toprule
\textbf{Dataset source}  & \textbf{Relevant studies}                              & \textbf{Count} \\ \midrule
GitHub & PS3, PS4, PS6, PS7, PS8, PS9, PS11, PS12, PS17, PS18, PS21, PS22, PS28, PS29, PS30, PS31, PS33, PS34, PS35, PS36, PS40, PS42, PS45, PS46  & 26   
\\ \midrule
Bugzilla & PS1, PS2, PS5, PS6, PS9, PS15, PS16, PS19, PS23, PS25,  PS37, PS38, PS39, PS41, PS43, PS44  & 17 \\\midrule
Jira & PS2, PS6, PS13, PS19, PS20, PS23, PS37, PS39  & 8  \\ \midrule
CSS data & PS10, PS26, PS32  & 3  \\ \midrule

Mantis, Redhat Bugzilla   & PS24 & 1 \\ \midrule
Twitter, Other forum posts  & PS27 & 1 \\ \bottomrule
\label{tab:data-sources}
\end{tabular}
\end{table}
 
\begin{table}[!ht]
\centering
\caption{An excerpt of the mapping of automatic techniques for issue report classification. Precision (P), Recall (R), F-score (F1), Accuracy (Acc), Micro Average (Avg), Test on a fixed set (TFS), and 10-fold CV (10FCV)}
\label{tab:excerpt-map} % 11.8 cm table
\begin{tabular}{p{0.5cm}p{2.5cm}p{1cm}p{1.5cm}p{1.5cm}p{1.5cm}p{1.5cm}}
\toprule
  \textbf{PS} & \textbf{Technique} & \textbf{Data source} & \textbf{Features} & \textbf{Pre-processing technique} & \textbf{Evaluation strategy} & \textbf{Evaluation metrics}
\\ \midrule

PS1 & Rule-based technique & Bugzilla & Title, description & TF-IDF & TFS & P, R \\ \midrule

PS2 & Random forest, SVM, K-means clustering  & Bugzilla, Jira & Summary & TF, TF-IDF, TF-IGM & TFS & R, P, F1 \\ \midrule

PS3 & BERT-based model, RoBERTa, CodeBERT, BERTOverflow,  seBERT & GitHub & Title, description & BERT-based tokenizer & TFS & Micro F1 \\ \midrule

PS4 & CNN & GitHub & Image, code, & Embedding layer & TFS & P, R, F1 \\ \midrule

%PS5 & KNN & Bugzilla, Redhat Bugzilla & Summary, severity, priority, component, assignee, reporter, OS, reproducibility & TF-IDF & TFS & R, P, F1 \\ \midrule

PS6 & Stacking ensemble, Random forest, SVM, Logistic regression, Multilayer perceptron, SGD, Multinomial Naive Bayes, Naive Bayes & GitHub, Bugzilla, Jira & Title, description & TF-IDF & Stratified 10FCV & Acc, MCC, F1 \\ \midrule

PS7 & SGD & GitHub & Title, body & TF-IDF & TFS & P, R,  F1,  Avg P, Avg R , Avg F1 \\ \midrule

PS8 & GPT-3.5-turbo base model & GitHub & Title, body & GPT-based tokenizer & TFS & P, R,  F1,  Avg P, Avg R , Avg F1 \\ \bottomrule

PS9 & Random tree, SVM, Naive Bayes & GitHub, Bugzilla & Title, description & Manual selection of textual features & 5-fold CV & Acc \\ \midrule

PS10 & Naive Bayes, SVM, KNN, Logistic regression, Decision tree, Random forest & CSS data & Description & TF-IDF & N-fold CV & F1 \\ \midrule

PS11 & RoBERTa & GitHub & Title, body & RoBERTa-based tokenizer & TFS & P, R, F1 \\ \midrule

PS12 & BERT-based model & GitHub & Title, body & BERT-based tokenizer & TFS & P, R,  F1,  Avg P, Avg R , Avg F1 \\ \bottomrule
        
\end{tabular}
\end{table}

\section{Discussion}\label{sec:discussion}

In the following subsections, we discuss (a) key lessons learned from this mapping study on issue report classification and (b) potential areas for future research.

\subsection{Prediction accuracy of automatic techniques for issue report classification}
The results of this mapping study show that the existing literature on issue report classification has evaluated various automatic techniques, including traditional ML and deep learning-based techniques, as well as more advanced large language models. 
Among these techniques, BERT-based models and their variants have demonstrated the highest performance [PS32, PS34, PS46]. In particular, RoBERTa has emerged as the state-of-the-art model within the BERT family, as reported by Izadi [PS46]. They evaluated RoBERTa using more than 1 million issue reports from GitHub projects \cite{nlbse2023} and found that it can classify these issue reports with a micro-average F1 score of 0.89 [PS7, PS46].

Furthermore, the findings from the literature indicate that traditional ML techniques such as logistic regression, SGD-based classifier, and SVM can achieve performance comparable to those complex deep learning techniques and BERT-based models [PS7, PS16, PS32]. For example, Laiq [PS7] trained an SGD-based classifier on more than 1 million issue reports from GitHub projects \cite{nlbse2023} and found that it can classify issue reports with a micro-average F1 score of 0.85.

\vspace{\baselineskip}
\noindent
\fbox{\begin{minipage}[!]{0.98\columnwidth}%
\textbf{Lessons learned:}
\begin{enumerate}

\item BERT-based models (e.g., RoBERta) are state-of-the-art models for issue report classification [PS32, PS34, PS46].

\item Traditional ML techniques, such as logistic regression, SGD-based classifier, and SVM, can achieve prediction performance comparable to complex deep learning techniques and BERT-based models for the classification of issues report [PS7, PS16, PS32].

\end{enumerate}
\end{minipage}}
%\vspace{\baselineskip}

\subsection{Lack of consideration of potentially relevant adoption factors}
We observe that in the existing literature on issue report classification, there is generally a lack of consideration of other potentially relevant adoption factors beyond prediction accuracy. Other factors of concern for practical adoption are, for example, maintenance cost, scalability, explainability, and adaptability \cite{rana2014framework,paleyes2022challenges}. These factors do not always complement each other and may require a trade-off. For example, while one technique may be \textit{more accurate} than another, it may score lower in terms of \textit{explainability}.

Approximately 89\% of the studies (41 out of 46) evaluated the proposed techniques solely based on prediction accuracy. Only five studies took into account the factors beyond prediction accuracy that may influence the adoption of these techniques in practice. Among those, three focused exclusively on explainability; however, their approach was limited and lacked real evaluation. Only two studies [PS26, PS32] thoroughly studied the adoption factors for ML techniques. These two studies emphasize that practitioners view these factors as important when deciding whether to adopt an ML technique in their practices. These findings also align with the existing literature that investigates factors influencing the adoption of ML techniques in practice \cite{rana2014framework, paleyes2022challenges}.

\vspace{\baselineskip}
\noindent
\fbox{\begin{minipage}[!]{0.98\columnwidth}%
\textbf{Lesson learned and call for future work:}
\begin{enumerate}%[leftmargin=*, topsep=0pt, itemsep=-1pt]

\item [] In the existing literature on issue report classification, there is generally a lack of consideration of potentially relevant other adoption factors beyond prediction accuracy. Therefore, future research should: a) take these factors into account when investigating issue report classification within a specific context and b) investigate how factors such as explainability, scalability, and generalizability, beyond just prediction accuracy, affect the practical adoption of ML techniques for automatic issue report classification.

\end{enumerate}
\end{minipage}}
%\vspace{\baselineskip}

\subsection{Lack of industrial validations and involvement of practitioners}
We observe that existing research on issue report classification primarily uses archival data from open-source repositories. Approximately 93\% (43/46) of the studies used archival
data. In contrast, only a small number, 3 out of 46, have used proprietary or industrial data. It is important to evaluate these techniques in real industrial settings to assess their performance, generalizability in other contexts, and practical applicability.

Furthermore, the involvement of practitioners in these studies is notably scarce. Approximately 91\% (42/46) of the studies did not involve practitioners. The involvement of practitioners is essential for gathering feedback on the proposed solutions regarding their practical relevance and applicability.

\vspace{\baselineskip}
\noindent
\fbox{\begin{minipage}[!]{0.98\columnwidth}%
\textbf{Lesson learned and call for future work:}
\begin{enumerate}%[leftmargin=*, topsep=0pt, itemsep=-1pt]

\item [] In the existing literature on issue report classification, in general, there is a lack of industrial validations and involvement of practitioners. Therefore, future work should consider the involvement of practitioners and the evaluation of these techniques in real industrial settings.

\end{enumerate}
\end{minipage}}
%\vspace{\baselineskip}

\subsection{Need to use robust evaluation metrics and approaches}
The data from the issue reports is often unbalanced. Therefore, it is important to use evaluation metrics that are robust to data imbalance, such as AUC and MCC, when assessing automatic techniques. However, we have observed a lack of these metrics in the literature. Only 6 of 46 studies have used AUC and only 5 of 46 have used MCC.

Similarly, we observe that robust evaluation approaches, such as ten-fold cross-validation, are lacking in the reviewed studies. The majority (29 out of 46) rely solely on a fixed test set for performance evaluation. In contrast, only 17 studies employ cross-validation techniques, which are generally considered more reliable to assess the performance of ML techniques.

\vspace{\baselineskip}
\noindent
\fbox{\begin{minipage}[!]{0.98\columnwidth}%
\textbf{Lesson learned and call for future work:}
\begin{enumerate}%[leftmargin=*, topsep=0pt, itemsep=-1pt]

\item [] In the existing literature on issue report classification, there is a lack of the use of robust evaluation metrics and approaches. Thus, future work on issue report classification should consider using more robust evaluation metrics and evaluation approaches such as AUC, MCC, and 10-fold CV.

\end{enumerate}
\end{minipage}}
%\vspace{\baselineskip}

\subsection{Reliability of ground truth used for issue report classification} 
An important but often overlooked issue in software engineering research is the reliability of ground truth (training data) used to train ML models \cite{tuzun2021ground}. This concern is equally relevant for automatic techniques used for issue report classification. Accurate labeling of issue reports, such as identifying them as bugs, features, questions, or documentation tasks, is essential for both training and evaluating ML models. Several studies \cite{herzig2013s,wu2021data,colavito2023few} have reported that these labels can be noisy, inconsistent, or incorrect. Herzig et al. \cite{herzig2013s} manually reviewed more than 7k issue reports and reported that approximately 33.8\% of the reports labeled as bugs were actually misclassified, including features, improvements, and documentation tasks. Similarly, Colvito et al. \cite{colavito2022issue} conducted an error analysis of the issue reports data \cite{nlbse2023} and reported the presence of noisy labels. ML models trained on such data will produce biased results and can lead to suboptimal decisions in practice. 

To address the above-mentioned issue, Colavito et al. \cite{colavito2022issue} constructed a new small data set by manually reviewing issue reports involving multiple reviewers. In the next step, they evaluated a few-shot learning-based framework, SETFIT, on the small data set and reported that it could improve performance compared to the models trained on noisy labels. These findings emphasize the effectiveness of incorporating small, high-quality labeled datasets to overcome labeling inconsistencies. Despite the proven impact of noisy ground truth on model accuracy and validity, very few studies explicitly address the quality of issue reports classification data in the current literature.

\vspace{\baselineskip}
\noindent
\fbox{\begin{minipage}[!]{0.98\columnwidth}%
\textbf{Lessons learned and call for future work:}
\begin{enumerate}%[leftmargin=*, topsep=0pt, itemsep=-1pt]

\item Issue report classification data (ground truth) can be noisy. Therefore, future work on issue report classification should carefully consider the reliability of the data for training and evaluating ML models.

\item Few-shot learning-based approaches (e.g., SetFIT) can help tackle the reliability issue of ground truth data, i.e., they can effectively classify issue reports using relatively small manually created reliable data sets. 

\end{enumerate}
\end{minipage}}
\vspace{\baselineskip}

\section{Conclusion}\label{sec:conclusion}

In this study, we provide a comprehensive map of the automatic techniques for issue report classification. Our results indicate that the existing literature applies various automatic techniques for issue report classification, including traditional ML and deep learning-based techniques and more advanced large language models. As reported in the literature, BERT-based models (e.g., RoBERta) are state-of-the-art models for issue report classification. In addition, traditional ML techniques, such as logistic regression, SGD-based classifier, and SVM, can achieve prediction performance comparable to complex deep learning techniques and BERT-based models for the classification of issue reports.

We found that these studies lack the involvement of practitioners and do not consider other potentially relevant adoption factors beyond prediction accuracy, such as the explainability, scalability, and generalizability of the techniques. The literature mainly uses archival data from open-source repositories. Therefore, future research should focus on real industrial settings, consider other potentially relevant adoption factors, and actively involve practitioners. In addition, it is essential for both researchers and practitioners to consider the lessons learned from the existing literature presented in this study when investigating the task of issue report classification within a specific context.

\section*{Statements and declarations}

\subsection*{Conflict of interest}
The authors declared that they have no conflict of interest.

\section*{Data availability} \label{sec:data-availability}
The data pertaining to this study can be found at (1) \url{https://tinyurl.com/wdkdxj65} and (2) \url{https://tinyurl.com/j92hvesa}.

% Link to map:
% Link to excluded studies:

%\section*{Acknowledgment}

% This work has been supported by XYZ...

%\section*{Author contributions}

\bibliography{sn-bibliography}


\begin{thebibliography}{70}
\ifx \bisbn   \undefined \def \bisbn  #1{ISBN #1}\fi
\ifx \binits  \undefined \def \binits#1{#1}\fi
\ifx \bauthor  \undefined \def \bauthor#1{#1}\fi
\ifx \batitle  \undefined \def \batitle#1{#1}\fi
\ifx \bjtitle  \undefined \def \bjtitle#1{#1}\fi
\ifx \bvolume  \undefined \def \bvolume#1{\textbf{#1}}\fi
\ifx \byear  \undefined \def \byear#1{#1}\fi
\ifx \bissue  \undefined \def \bissue#1{#1}\fi
\ifx \bfpage  \undefined \def \bfpage#1{#1}\fi
\ifx \blpage  \undefined \def \blpage #1{#1}\fi
\ifx \burl  \undefined \def \burl#1{\textsf{#1}}\fi
\ifx \doiurl  \undefined \def \doiurl#1{\url{https://doi.org/#1}}\fi
\ifx \betal  \undefined \def \betal{\textit{et al.}}\fi
\ifx \binstitute  \undefined \def \binstitute#1{#1}\fi
\ifx \binstitutionaled  \undefined \def \binstitutionaled#1{#1}\fi
\ifx \bctitle  \undefined \def \bctitle#1{#1}\fi
\ifx \beditor  \undefined \def \beditor#1{#1}\fi
\ifx \bpublisher  \undefined \def \bpublisher#1{#1}\fi
\ifx \bbtitle  \undefined \def \bbtitle#1{#1}\fi
\ifx \bedition  \undefined \def \bedition#1{#1}\fi
\ifx \bseriesno  \undefined \def \bseriesno#1{#1}\fi
\ifx \blocation  \undefined \def \blocation#1{#1}\fi
\ifx \bsertitle  \undefined \def \bsertitle#1{#1}\fi
\ifx \bsnm \undefined \def \bsnm#1{#1}\fi
\ifx \bsuffix \undefined \def \bsuffix#1{#1}\fi
\ifx \bparticle \undefined \def \bparticle#1{#1}\fi
\ifx \barticle \undefined \def \barticle#1{#1}\fi
\bibcommenthead
\ifx \bconfdate \undefined \def \bconfdate #1{#1}\fi
\ifx \botherref \undefined \def \botherref #1{#1}\fi
\ifx \url \undefined \def \url#1{\textsf{#1}}\fi
\ifx \bchapter \undefined \def \bchapter#1{#1}\fi
\ifx \bbook \undefined \def \bbook#1{#1}\fi
\ifx \bcomment \undefined \def \bcomment#1{#1}\fi
\ifx \oauthor \undefined \def \oauthor#1{#1}\fi
\ifx \citeauthoryear \undefined \def \citeauthoryear#1{#1}\fi
\ifx \endbibitem  \undefined \def \endbibitem {}\fi
\ifx \bconflocation  \undefined \def \bconflocation#1{#1}\fi
\ifx \arxivurl  \undefined \def \arxivurl#1{\textsf{#1}}\fi
\csname PreBibitemsHook\endcsname

%%% 1
\bibitem[\protect\citeauthoryear{Antoniol et~al.}{2008}]{antoniol2008bug}
\begin{bchapter}
\bauthor{\bsnm{Antoniol}, \binits{G.}},
\bauthor{\bsnm{Ayari}, \binits{K.}},
\bauthor{\bsnm{Di~Penta}, \binits{M.}},
\bauthor{\bsnm{Khomh}, \binits{F.}},
\bauthor{\bsnm{Gu{\'e}h{\'e}neuc}, \binits{Y.-G.}}:
\bctitle{Is it a bug or an enhancement? a text-based approach to classify change requests}.
In: \bbtitle{Proceedings of the 2008 Conference of the Center for Advanced Studies on Collaborative Research: Meeting of Minds},
pp. \bfpage{304}--\blpage{318}
(\byear{2008})
\end{bchapter}
\endbibitem

%%% 2
\bibitem[\protect\citeauthoryear{Herzig et~al.}{2013}]{herzig2013s}
\begin{bchapter}
\bauthor{\bsnm{Herzig}, \binits{K.}},
\bauthor{\bsnm{Just}, \binits{S.}},
\bauthor{\bsnm{Zeller}, \binits{A.}}:
\bctitle{It's not a bug, it's a feature: how misclassification impacts bug prediction}.
In: \bbtitle{2013 35th International Conference on Software Engineering (ICSE)},
pp. \bfpage{392}--\blpage{401}
(\byear{2013}).
\bcomment{IEEE}
\end{bchapter}
\endbibitem

%%% 3
\bibitem[\protect\citeauthoryear{Zanetti et~al.}{2013}]{zanetti2013categorizing}
\begin{bchapter}
\bauthor{\bsnm{Zanetti}, \binits{M.S.}},
\bauthor{\bsnm{Scholtes}, \binits{I.}},
\bauthor{\bsnm{Tessone}, \binits{C.J.}},
\bauthor{\bsnm{Schweitzer}, \binits{F.}}:
\bctitle{Categorizing bugs with social networks: a case study on four open source software communities}.
In: \bbtitle{2013 35th International Conference on Software Engineering (ICSE)},
pp. \bfpage{1032}--\blpage{1041}
(\byear{2013}).
\bcomment{IEEE}
\end{bchapter}
\endbibitem

%%% 4
\bibitem[\protect\citeauthoryear{Laiq et~al.}{2022}]{laiq2022early}
\begin{bchapter}
\bauthor{\bsnm{Laiq}, \binits{M.}},
\bauthor{\bsnm{Ali}, \binits{N.b.}},
\bauthor{\bsnm{B{\"o}stler}, \binits{J.}},
\bauthor{\bsnm{Engstr{\"o}m}, \binits{E.}}:
\bctitle{Early identification of invalid bug reports in industrial settings--a case study}.
In: \bbtitle{International Conference on Product-Focused Software Process Improvement},
pp. \bfpage{497}--\blpage{507}
(\byear{2022}).
\bcomment{Springer}
\end{bchapter}
\endbibitem

%%% 5
\bibitem[\protect\citeauthoryear{Cho et~al.}{2022}]{cho2022classifying}
\begin{barticle}
\bauthor{\bsnm{Cho}, \binits{H.}},
\bauthor{\bsnm{Lee}, \binits{S.}},
\bauthor{\bsnm{Kang}, \binits{S.}}:
\batitle{Classifying issue reports according to feature descriptions in a user manual based on a deep learning model}.
\bjtitle{Information and Software Technology}
\bvolume{142},
\bfpage{106743}
(\byear{2022})
\end{barticle}
\endbibitem

%%% 6
\bibitem[\protect\citeauthoryear{Siddiq and Santos}{2022}]{siddiq2022bert}
\begin{bchapter}
\bauthor{\bsnm{Siddiq}, \binits{M.L.}},
\bauthor{\bsnm{Santos}, \binits{J.C.}}:
\bctitle{Bert-based github issue report classification}.
In: \bbtitle{Proceedings of the 1st International Workshop on Natural Language-based Software Engineering},
pp. \bfpage{33}--\blpage{36}
(\byear{2022})
\end{bchapter}
\endbibitem

%%% 7
\bibitem[\protect\citeauthoryear{Petersen et~al.}{2008}]{petersen2008systematic}
\begin{bchapter}
\bauthor{\bsnm{Petersen}, \binits{K.}},
\bauthor{\bsnm{Feldt}, \binits{R.}},
\bauthor{\bsnm{Mujtaba}, \binits{S.}},
\bauthor{\bsnm{Mattsson}, \binits{M.}}:
\bctitle{Systematic mapping studies in software engineering}.
In: \bbtitle{International Conference on Evaluation and Assessment in Software Engineering},
pp. \bfpage{1}--\blpage{10}
(\byear{2008})
\end{bchapter}
\endbibitem

%%% 8
\bibitem[\protect\citeauthoryear{Zou et~al.}{2018}]{zou2018practitioners}
\begin{barticle}
\bauthor{\bsnm{Zou}, \binits{W.}},
\bauthor{\bsnm{Lo}, \binits{D.}},
\bauthor{\bsnm{Chen}, \binits{Z.}},
\bauthor{\bsnm{Xia}, \binits{X.}},
\bauthor{\bsnm{Feng}, \binits{Y.}},
\bauthor{\bsnm{Xu}, \binits{B.}}:
\batitle{How practitioners perceive automated bug report management techniques}.
\bjtitle{IEEE Transactions on Software Engineering}
\bvolume{46}(\bissue{8}),
\bfpage{836}--\blpage{862}
(\byear{2018})
\end{barticle}
\endbibitem

%%% 9
\bibitem[\protect\citeauthoryear{Parnin and Orso}{2011}]{parnin2011automated}
\begin{bchapter}
\bauthor{\bsnm{Parnin}, \binits{C.}},
\bauthor{\bsnm{Orso}, \binits{A.}}:
\bctitle{Are automated debugging techniques actually helping programmers?}
In: \bbtitle{Proceedings of the 2011 International Symposium on Software Testing and Analysis},
pp. \bfpage{199}--\blpage{209}
(\byear{2011})
\end{bchapter}
\endbibitem

%%% 10
\bibitem[\protect\citeauthoryear{Laiq et~al.}{2023}]{laiq2023data}
\begin{barticle}
\bauthor{\bsnm{Laiq}, \binits{M.}},
\bauthor{\bsnm{Ali}, \binits{N.}},
\bauthor{\bsnm{B{\"o}rstler}, \binits{J.}},
\bauthor{\bsnm{Engstr{\"o}m}, \binits{E.}}:
\batitle{A data-driven approach for understanding invalid bug reports: An industrial case study}.
\bjtitle{Information and Software Technology}
\bvolume{164},
\bfpage{107305}
(\byear{2023})
\end{barticle}
\endbibitem

%%% 11
\bibitem[\protect\citeauthoryear{Fan et~al.}{2018}]{fan2018chaff}
\begin{barticle}
\bauthor{\bsnm{Fan}, \binits{Y.}},
\bauthor{\bsnm{Xia}, \binits{X.}},
\bauthor{\bsnm{Lo}, \binits{D.}},
\bauthor{\bsnm{Hassan}, \binits{A.E.}}:
\batitle{Chaff from the wheat: Characterizing and determining valid bug reports}.
\bjtitle{IEEE transactions on software engineering}
\bvolume{46}(\bissue{5}),
\bfpage{495}--\blpage{525}
(\byear{2018})
\end{barticle}
\endbibitem

%%% 12
\bibitem[\protect\citeauthoryear{Zhang et~al.}{2015}]{zhang2015survey}
\begin{barticle}
\bauthor{\bsnm{Zhang}, \binits{J.}},
\bauthor{\bsnm{Wang}, \binits{X.}},
\bauthor{\bsnm{Hao}, \binits{D.}},
\bauthor{\bsnm{Xie}, \binits{B.}},
\bauthor{\bsnm{Zhang}, \binits{L.}},
\bauthor{\bsnm{Mei}, \binits{H.}}:
\batitle{A survey on bug-report analysis}.
\bjtitle{Sci. China Inf. Sci.}
\bvolume{58}(\bissue{2}),
\bfpage{1}--\blpage{24}
(\byear{2015})
\end{barticle}
\endbibitem

%%% 13
\bibitem[\protect\citeauthoryear{Borg et~al.}{2024}]{borg2024adopting}
\begin{barticle}
\bauthor{\bsnm{Borg}, \binits{M.}},
\bauthor{\bsnm{Jonsson}, \binits{L.}},
\bauthor{\bsnm{Engstr{\"o}m}, \binits{E.}},
\bauthor{\bsnm{Bartalos}, \binits{B.}},
\bauthor{\bsnm{Szab{\'o}}, \binits{A.}}:
\batitle{Adopting automated bug assignment in practice—a longitudinal case study at ericsson}.
\bjtitle{Empirical Software Engineering}
\bvolume{29}(\bissue{5}),
\bfpage{126}
(\byear{2024})
\end{barticle}
\endbibitem

%%% 14
\bibitem[\protect\citeauthoryear{Kim et~al.}{2013}]{kim2013should}
\begin{barticle}
\bauthor{\bsnm{Kim}, \binits{D.}},
\bauthor{\bsnm{Tao}, \binits{Y.}},
\bauthor{\bsnm{Kim}, \binits{S.}},
\bauthor{\bsnm{Zeller}, \binits{A.}}:
\batitle{Where should we fix this bug? a two-phase recommendation model}.
\bjtitle{IEEE transactions on software Engineering}
\bvolume{39}(\bissue{11}),
\bfpage{1597}--\blpage{1610}
(\byear{2013})
\end{barticle}
\endbibitem

%%% 15
\bibitem[\protect\citeauthoryear{Zhang et~al.}{2023}]{zhang2023duplicate}
\begin{barticle}
\bauthor{\bsnm{Zhang}, \binits{T.}},
\bauthor{\bsnm{Han}, \binits{D.}},
\bauthor{\bsnm{Vinayakarao}, \binits{V.}},
\bauthor{\bsnm{Irsan}, \binits{I.C.}},
\bauthor{\bsnm{Xu}, \binits{B.}},
\bauthor{\bsnm{Thung}, \binits{F.}},
\bauthor{\bsnm{Lo}, \binits{D.}},
\bauthor{\bsnm{Jiang}, \binits{L.}}:
\batitle{Duplicate bug report detection: How far are we?}
\bjtitle{ACM Transactions on Software Engineering and Methodology}
\bvolume{32}(\bissue{4}),
\bfpage{1}--\blpage{32}
(\byear{2023})
\end{barticle}
\endbibitem

%%% 16
\bibitem[\protect\citeauthoryear{Colavito et~al.}{2024}]{colavito2024large}
\begin{botherref}
\oauthor{\bsnm{Colavito}, \binits{G.}},
\oauthor{\bsnm{Lanubile}, \binits{F.}},
\oauthor{\bsnm{Novielli}, \binits{N.}},
\oauthor{\bsnm{Quaranta}, \binits{L.}}:
Large language models for issue report classification
(2024)
\end{botherref}
\endbibitem

%%% 17
\bibitem[\protect\citeauthoryear{He et~al.}{2020}]{he2020deep}
\begin{bchapter}
\bauthor{\bsnm{He}, \binits{J.}},
\bauthor{\bsnm{Xu}, \binits{L.}},
\bauthor{\bsnm{Fan}, \binits{Y.}},
\bauthor{\bsnm{Xu}, \binits{Z.}},
\bauthor{\bsnm{Yan}, \binits{M.}},
\bauthor{\bsnm{Lei}, \binits{Y.}}:
\bctitle{Deep learning based valid bug reports determination and explanation}.
In: \bbtitle{2020 IEEE 31st International Symposium on Software Reliability Engineering (ISSRE)},
pp. \bfpage{184}--\blpage{194}
(\byear{2020}).
\bcomment{IEEE}
\end{bchapter}
\endbibitem

%%% 18
\bibitem[\protect\citeauthoryear{Kwak et~al.}{2023}]{kwak2023multimodal}
\begin{barticle}
\bauthor{\bsnm{Kwak}, \binits{C.}},
\bauthor{\bsnm{Jung}, \binits{P.}},
\bauthor{\bsnm{Lee}, \binits{S.}}:
\batitle{A multimodal deep learning model using text, image, and code data for improving issue classification tasks}.
\bjtitle{Applied Sciences}
\bvolume{13}(\bissue{16}),
\bfpage{9456}
(\byear{2023})
\end{barticle}
\endbibitem

%%% 19
\bibitem[\protect\citeauthoryear{Heo et~al.}{2024}]{heo2024comparison}
\begin{botherref}
\oauthor{\bsnm{Heo}, \binits{J.}},
\oauthor{\bsnm{Kwon}, \binits{G.}},
\oauthor{\bsnm{Kwak}, \binits{C.}},
\oauthor{\bsnm{Lee}, \binits{S.}}:
A comparison of pretrained models for classifying issue reports.
IEEE Access
(2024)
\end{botherref}
\endbibitem

%%% 20
\bibitem[\protect\citeauthoryear{Laiq et~al.}{2024}]{laiq2024industrial}
\begin{barticle}
\bauthor{\bsnm{Laiq}, \binits{M.}},
\bauthor{\bsnm{Ali}, \binits{N.b.}},
\bauthor{\bsnm{B{\"o}rstler}, \binits{J.}},
\bauthor{\bsnm{Engstr{\"o}m}, \binits{E.}}:
\batitle{Industrial adoption of machine learning techniques for early identification of invalid bug reports}.
\bjtitle{Empirical Software Engineering}
\bvolume{29}(\bissue{5}),
\bfpage{130}
(\byear{2024})
\end{barticle}
\endbibitem

%%% 21
\bibitem[\protect\citeauthoryear{K{\"o}ksal and {\"O}zt{\"u}rk}{2022}]{koksal2022survey}
\begin{bchapter}
\bauthor{\bsnm{K{\"o}ksal}, \binits{{\"O}.}},
\bauthor{\bsnm{{\"O}zt{\"u}rk}, \binits{C.E.}}:
\bctitle{A survey on machine learning-based automated software bug report classification}.
In: \bbtitle{2022 International Symposium on Multidisciplinary Studies and Innovative Technologies (ISMSIT)},
pp. \bfpage{635}--\blpage{640}
(\byear{2022}).
\bcomment{IEEE}
\end{bchapter}
\endbibitem

%%% 22
\bibitem[\protect\citeauthoryear{Kitchenham et~al.}{2016}]{kitchenham2016evidence}
\begin{bbook}
\bauthor{\bsnm{Kitchenham}, \binits{B.A.}},
\bauthor{\bsnm{Budgen}, \binits{D.}},
\bauthor{\bsnm{Brereton}, \binits{P.}}:
\bbtitle{Evidence-based Software Engineering and Systematic Reviews}.
\bpublisher{CRC press}, \blocation{???}
(\byear{2016})
\end{bbook}
\endbibitem

%%% 23
\bibitem[\protect\citeauthoryear{Gomes et~al.}{2019}]{gomes2019bug}
\begin{barticle}
\bauthor{\bsnm{Gomes}, \binits{L.A.F.}},
\bauthor{\bsnm{Silva~Torres}, \binits{R.}},
\bauthor{\bsnm{C{\^o}rtes}, \binits{M.L.}}:
\batitle{Bug report severity level prediction in open source software: A survey and research opportunities}.
\bjtitle{Information and software technology}
\bvolume{115},
\bfpage{58}--\blpage{78}
(\byear{2019})
\end{barticle}
\endbibitem

%%% 24
\bibitem[\protect\citeauthoryear{Uddin et~al.}{2017}]{uddin2017survey}
\begin{barticle}
\bauthor{\bsnm{Uddin}, \binits{J.}},
\bauthor{\bsnm{Ghazali}, \binits{R.}},
\bauthor{\bsnm{Deris}, \binits{M.M.}},
\bauthor{\bsnm{Naseem}, \binits{R.}},
\bauthor{\bsnm{Shah}, \binits{H.}}:
\batitle{A survey on bug prioritization}.
\bjtitle{Artificial Intelligence Review}
\bvolume{47},
\bfpage{145}--\blpage{180}
(\byear{2017})
\end{barticle}
\endbibitem

%%% 25
\bibitem[\protect\citeauthoryear{Ampatzoglou et~al.}{2019}]{ampatzoglou2019identifying}
\begin{barticle}
\bauthor{\bsnm{Ampatzoglou}, \binits{A.}},
\bauthor{\bsnm{Bibi}, \binits{S.}},
\bauthor{\bsnm{Avgeriou}, \binits{P.}},
\bauthor{\bsnm{Verbeek}, \binits{M.}},
\bauthor{\bsnm{Chatzigeorgiou}, \binits{A.}}:
\batitle{Identifying, categorizing and mitigating threats to validity in software engineering secondary studies}.
\bjtitle{Information and Software Technology}
\bvolume{106},
\bfpage{201}--\blpage{230}
(\byear{2019})
\doiurl{10.1016/j.infsof.2018.10.006}
\end{barticle}
\endbibitem

%%% 26
\bibitem[\protect\citeauthoryear{Usman et~al.}{2023}]{usman2023quality}
\begin{barticle}
\bauthor{\bsnm{Usman}, \binits{M.}},
\bauthor{\bsnm{Ali}, \binits{N.B.}},
\bauthor{\bsnm{Wohlin}, \binits{C.}}:
\batitle{A quality assessment instrument for systematic literature reviews in software engineering}.
\bjtitle{e-Informatica Software Engineering Journal}
\bvolume{17}(\bissue{1}),
\bfpage{230105}
(\byear{2023})
\doiurl{10.37190/E-INF230105}
\end{barticle}
\endbibitem

%%% 27
\bibitem[\protect\citeauthoryear{Kitchenham}{2010}]{kitchenham2010s}
\begin{barticle}
\bauthor{\bsnm{Kitchenham}, \binits{B.}}:
\batitle{What’s up with software metrics?--a preliminary mapping study}.
\bjtitle{Journal of systems and software}
\bvolume{83}(\bissue{1}),
\bfpage{37}--\blpage{51}
(\byear{2010})
\end{barticle}
\endbibitem

%%% 28
\bibitem[\protect\citeauthoryear{Petersen}{2011}]{petersen2011measuring}
\begin{barticle}
\bauthor{\bsnm{Petersen}, \binits{K.}}:
\batitle{Measuring and predicting software productivity: A systematic map and review}.
\bjtitle{Information and Software Technology}
\bvolume{53}(\bissue{4}),
\bfpage{317}--\blpage{343}
(\byear{2011})
\end{barticle}
\endbibitem

%%% 29
\bibitem[\protect\citeauthoryear{Ali}{2016}]{Ali16}
\begin{bchapter}
\bauthor{\bsnm{Ali}, \binits{N.B.}}:
\bctitle{Is effectiveness sufficient to choose an intervention?: Considering resource use in empirical software engineering}.
In: \bbtitle{International Symposium on Empirical Software Engineering and Measurement},
pp. \bfpage{54}--\blpage{1546}.
\bpublisher{{ACM}}, \blocation{???}
(\byear{2016}).
\doiurl{10.1145/2961111.2962631} .
\burl{https://doi.org/10.1145/2961111.2962631}
\end{bchapter}
\endbibitem

%%% 30
\bibitem[\protect\citeauthoryear{Rana et~al.}{2014}]{rana2014framework}
\begin{bchapter}
\bauthor{\bsnm{Rana}, \binits{R.}},
\bauthor{\bsnm{Staron}, \binits{M.}},
\bauthor{\bsnm{Hansson}, \binits{J.}},
\bauthor{\bsnm{Nilsson}, \binits{M.}},
\bauthor{\bsnm{Meding}, \binits{W.}}:
\bctitle{A framework for adoption of machine learning in industry for software defect prediction}.
In: \bbtitle{Proceedings of the 9th International Conference on Software Engineering and Applications},
pp. \bfpage{383}--\blpage{392}
(\byear{2014})
\end{bchapter}
\endbibitem

%%% 31
\bibitem[\protect\citeauthoryear{Paleyes et~al.}{2022}]{paleyes2022challenges}
\begin{barticle}
\bauthor{\bsnm{Paleyes}, \binits{A.}},
\bauthor{\bsnm{Urma}, \binits{R.-G.}},
\bauthor{\bsnm{Lawrence}, \binits{N.D.}}:
\batitle{Challenges in deploying machine learning: {A} survey of case studies}.
\bjtitle{ACM Computing Surveys}
\bvolume{55}(\bissue{6}),
\bfpage{1}--\blpage{29}
(\byear{2022})
\end{barticle}
\endbibitem

%%% 32
\bibitem[\protect\citeauthoryear{Kallis et~al.}{2023}]{nlbse2023}
\begin{bchapter}
\bauthor{\bsnm{Kallis}, \binits{R.}},
\bauthor{\bsnm{Izadi}, \binits{M.}},
\bauthor{\bsnm{Pascarella}, \binits{L.}},
\bauthor{\bsnm{Chaparro}, \binits{O.}},
\bauthor{\bsnm{Rani}, \binits{P.}}:
\bctitle{The nlbse'23 tool competition}.
In: \bbtitle{Proceedings of The 2nd International Workshop on Natural Language-based Software Engineering (NLBSE'23)}
(\byear{2023})
\end{bchapter}
\endbibitem

%%% 33
\bibitem[\protect\citeauthoryear{T{\"u}z{\"u}n et~al.}{2021}]{tuzun2021ground}
\begin{barticle}
\bauthor{\bsnm{T{\"u}z{\"u}n}, \binits{E.}},
\bauthor{\bsnm{Erdogmus}, \binits{H.}},
\bauthor{\bsnm{Baldassarre}, \binits{M.T.}},
\bauthor{\bsnm{Felderer}, \binits{M.}},
\bauthor{\bsnm{Feldt}, \binits{R.}},
\bauthor{\bsnm{Turhan}, \binits{B.}}:
\batitle{Ground-truth deficiencies in software engineering: when codifying the past can be counterproductive}.
\bjtitle{IEEE Software}
\bvolume{39}(\bissue{3}),
\bfpage{85}--\blpage{95}
(\byear{2021})
\end{barticle}
\endbibitem

%%% 34
\bibitem[\protect\citeauthoryear{Wu et~al.}{2021}]{wu2021data}
\begin{barticle}
\bauthor{\bsnm{Wu}, \binits{X.}},
\bauthor{\bsnm{Zheng}, \binits{W.}},
\bauthor{\bsnm{Xia}, \binits{X.}},
\bauthor{\bsnm{Lo}, \binits{D.}}:
\batitle{Data quality matters: A case study on data label correctness for security bug report prediction}.
\bjtitle{IEEE Transactions on Software Engineering}
\bvolume{48}(\bissue{7}),
\bfpage{2541}--\blpage{2556}
(\byear{2021})
\end{barticle}
\endbibitem

%%% 35
\bibitem[\protect\citeauthoryear{Colavito et~al.}{2023}]{colavito2023few}
\begin{bchapter}
\bauthor{\bsnm{Colavito}, \binits{G.}},
\bauthor{\bsnm{Lanubile}, \binits{F.}},
\bauthor{\bsnm{Novielli}, \binits{N.}}:
\bctitle{Few-shot learning for issue report classification}.
In: \bbtitle{2023 IEEE/ACM 2nd International Workshop on Natural Language-Based Software Engineering (NLBSE)},
pp. \bfpage{16}--\blpage{19}
(\byear{2023}).
\bcomment{IEEE}
\end{bchapter}
\endbibitem

%%% 36
\bibitem[\protect\citeauthoryear{Colavito et~al.}{2022}]{colavito2022issue}
\begin{bchapter}
\bauthor{\bsnm{Colavito}, \binits{G.}},
\bauthor{\bsnm{Lanubile}, \binits{F.}},
\bauthor{\bsnm{Novielli}, \binits{N.}}:
\bctitle{Issue report classification using pre-trained language models}.
In: \bbtitle{Proceedings of the 1st International Workshop on Natural Language-based Software Engineering},
pp. \bfpage{29}--\blpage{32}
(\byear{2022})
\end{bchapter}
\endbibitem

%%% 37
\bibitem[\protect\citeauthoryear{Zhang and Lee}{2011}]{zhang2011bug}
\begin{bchapter}
\bauthor{\bsnm{Zhang}, \binits{T.}},
\bauthor{\bsnm{Lee}, \binits{B.}}:
\bctitle{A bug rule based technique with feedback for classifying bug reports}.
In: \bbtitle{2011 IEEE 11th International Conference on Computer and Information Technology},
pp. \bfpage{336}--\blpage{343}
(\byear{2011}).
\bcomment{IEEE}
\end{bchapter}
\endbibitem

%%% 38
\bibitem[\protect\citeauthoryear{Polpinij et~al.}{2022}]{polpinij2022comparative}
\begin{bchapter}
\bauthor{\bsnm{Polpinij}, \binits{J.}},
\bauthor{\bsnm{Kaenampornpan}, \binits{M.}},
\bauthor{\bsnm{Luaphol}, \binits{B.}}:
\bctitle{A comparative study of short text classification methods for bug report type identification}.
In: \bbtitle{2022 Research, Invention, and Innovation Congress: Innovative Electricals and Electronics (RI2C)},
pp. \bfpage{27}--\blpage{33}
(\byear{2022}).
\bcomment{IEEE}
\end{bchapter}
\endbibitem

%%% 39
\bibitem[\protect\citeauthoryear{Kukkar and Mohana}{2018}]{kukkar2018supervised}
\begin{barticle}
\bauthor{\bsnm{Kukkar}, \binits{A.}},
\bauthor{\bsnm{Mohana}, \binits{R.}}:
\batitle{A supervised bug report classification with incorporate and textual field knowledge}.
\bjtitle{Procedia computer science}
\bvolume{132},
\bfpage{352}--\blpage{361}
(\byear{2018})
\end{barticle}
\endbibitem

%%% 40
\bibitem[\protect\citeauthoryear{Alraddadi and Alshayeb}{2024}]{alraddadi2024empirical}
\begin{botherref}
\oauthor{\bsnm{Alraddadi}, \binits{R.}},
\oauthor{\bsnm{Alshayeb}, \binits{M.}}:
An empirical evaluation of stacked generalization models for binary bug report classification.
Innovations in Systems and Software Engineering,
1--16
(2024)
\end{botherref}
\endbibitem

%%% 41
\bibitem[\protect\citeauthoryear{Laiq}{2023}]{laiq2023intelligent}
\begin{bchapter}
\bauthor{\bsnm{Laiq}, \binits{M.}}:
\bctitle{An intelligent tool for classifying issue reports}.
In: \bbtitle{2023 IEEE/ACM 2nd International Workshop on Natural Language-Based Software Engineering (NLBSE)},
pp. \bfpage{13}--\blpage{15}
(\byear{2023}).
\bcomment{IEEE}
\end{bchapter}
\endbibitem

%%% 42
\bibitem[\protect\citeauthoryear{Aracena et~al.}{2024}]{aracena2024applying}
\begin{bchapter}
\bauthor{\bsnm{Aracena}, \binits{G.}},
\bauthor{\bsnm{Luster}, \binits{K.}},
\bauthor{\bsnm{Santos}, \binits{F.}},
\bauthor{\bsnm{Steinmacher}, \binits{I.}},
\bauthor{\bsnm{Gerosa}, \binits{M.A.}}:
\bctitle{Applying large language models to issue classification}.
In: \bbtitle{Proceedings of the Third ACM/IEEE International Workshop on NL-based Software Engineering},
pp. \bfpage{57}--\blpage{60}
(\byear{2024})
\end{bchapter}
\endbibitem

%%% 43
\bibitem[\protect\citeauthoryear{Otoom et~al.}{2019}]{otoom2019automated}
\begin{bchapter}
\bauthor{\bsnm{Otoom}, \binits{A.F.}},
\bauthor{\bsnm{Al-jdaeh}, \binits{S.}},
\bauthor{\bsnm{Hammad}, \binits{M.}}:
\bctitle{Automated classification of software bug reports}.
In: \bbtitle{Proceedings of the 9th International Conference on Information Communication and Management},
pp. \bfpage{17}--\blpage{21}
(\byear{2019})
\end{bchapter}
\endbibitem

%%% 44
\bibitem[\protect\citeauthoryear{K{\"o}ksal and Tekinerdogan}{2021}]{koksal2021automated}
\begin{barticle}
\bauthor{\bsnm{K{\"o}ksal}, \binits{{\"O}.}},
\bauthor{\bsnm{Tekinerdogan}, \binits{B.}}:
\batitle{Automated classification of unstructured bilingual software bug reports: An industrial case study research}.
\bjtitle{Applied Sciences}
\bvolume{12}(\bissue{1}),
\bfpage{338}
(\byear{2021})
\end{barticle}
\endbibitem

%%% 45
\bibitem[\protect\citeauthoryear{Nadeem et~al.}{2021}]{nadeem2021automatic}
\begin{bchapter}
\bauthor{\bsnm{Nadeem}, \binits{A.}},
\bauthor{\bsnm{Sarwar}, \binits{M.U.}},
\bauthor{\bsnm{Malik}, \binits{M.Z.}}:
\bctitle{Automatic issue classifier: A transfer learning framework for classifying issue reports}.
In: \bbtitle{2021 IEEE International Symposium on Software Reliability Engineering Workshops (ISSREW)},
pp. \bfpage{421}--\blpage{426}
(\byear{2021}).
\bcomment{IEEE}
\end{bchapter}
\endbibitem

%%% 46
\bibitem[\protect\citeauthoryear{Terdchanakul et~al.}{2017}]{terdchanakul2017bug}
\begin{bchapter}
\bauthor{\bsnm{Terdchanakul}, \binits{P.}},
\bauthor{\bsnm{Hata}, \binits{H.}},
\bauthor{\bsnm{Phannachitta}, \binits{P.}},
\bauthor{\bsnm{Matsumoto}, \binits{K.}}:
\bctitle{Bug or not? bug report classification using n-gram idf}.
In: \bbtitle{2017 IEEE International Conference on Software Maintenance and Evolution (ICSME)},
pp. \bfpage{534}--\blpage{538}
(\byear{2017}).
\bcomment{IEEE}
\end{bchapter}
\endbibitem

%%% 47
\bibitem[\protect\citeauthoryear{Chen et~al.}{2019}]{chen2019bug}
\begin{barticle}
\bauthor{\bsnm{Chen}, \binits{L.}},
\bauthor{\bsnm{Huang}, \binits{S.}},
\bauthor{\bsnm{Sun}, \binits{J.}},
\bauthor{\bsnm{Hui}, \binits{Z.}},
\bauthor{\bsnm{Yang}, \binits{S.}}:
\batitle{Bug report classification based on vector space model}.
\bjtitle{International Journal of Performability Engineering}
\bvolume{15}(\bissue{8}),
\bfpage{2071}
(\byear{2019})
\end{barticle}
\endbibitem

%%% 48
\bibitem[\protect\citeauthoryear{Alam et~al.}{2024}]{alam2024classifai}
\begin{bchapter}
\bauthor{\bsnm{Alam}, \binits{K.A.}},
\bauthor{\bsnm{Jumani}, \binits{A.}},
\bauthor{\bsnm{Aamir}, \binits{H.}},
\bauthor{\bsnm{Uzair}, \binits{M.}}:
\bctitle{Classifai: Automating issue reports classification using pre-trained bert (bidirectional encoder representations from transformers) models}.
In: \bbtitle{Proceedings of the Third ACM/IEEE International Workshop on NL-based Software Engineering},
pp. \bfpage{49}--\blpage{52}
(\byear{2024})
\end{bchapter}
\endbibitem

%%% 49
\bibitem[\protect\citeauthoryear{Zhifang et~al.}{2024}]{zhifang2024classification}
\begin{barticle}
\bauthor{\bsnm{Zhifang}, \binits{L.}},
\bauthor{\bsnm{Kun}, \binits{W.}},
\bauthor{\bsnm{Qi}, \binits{Z.}},
\bauthor{\bsnm{Shengzong}, \binits{L.}},
\bauthor{\bsnm{Yan}, \binits{Z.}},
\bauthor{\bsnm{Jianbiao}, \binits{H.}}:
\batitle{Classification of open source software bug report based on transfer learning}.
\bjtitle{Expert Systems}
\bvolume{41}(\bissue{5}),
\bfpage{13184}
(\byear{2024})
\end{barticle}
\endbibitem

%%% 50
\bibitem[\protect\citeauthoryear{Qin and Sun}{2018}]{qin2018classifying}
\begin{bchapter}
\bauthor{\bsnm{Qin}, \binits{H.}},
\bauthor{\bsnm{Sun}, \binits{X.}}:
\bctitle{Classifying bug reports into bugs and non-bugs using lstm}.
In: \bbtitle{Proceedings of the 10th Asia-Pacific Symposium on Internetware},
pp. \bfpage{1}--\blpage{4}
(\byear{2018})
\end{bchapter}
\endbibitem

%%% 51
\bibitem[\protect\citeauthoryear{Pingclasai et~al.}{2013}]{pingclasai2013classifying}
\begin{bchapter}
\bauthor{\bsnm{Pingclasai}, \binits{N.}},
\bauthor{\bsnm{Hata}, \binits{H.}},
\bauthor{\bsnm{Matsumoto}, \binits{K.-i.}}:
\bctitle{Classifying bug reports to bugs and other requests using topic modeling}.
In: \bbtitle{2013 20Th Asia-pacific Software Engineering Conference (APSEC)},
vol. \bseriesno{2},
pp. \bfpage{13}--\blpage{18}
(\byear{2013}).
\bcomment{IEEE}
\end{bchapter}
\endbibitem

%%% 52
\bibitem[\protect\citeauthoryear{Park et~al.}{2022}]{park2022classifying}
\begin{bchapter}
\bauthor{\bsnm{Park}, \binits{D.}},
\bauthor{\bsnm{Cho}, \binits{H.}},
\bauthor{\bsnm{Lee}, \binits{S.}}:
\bctitle{Classifying issues into custom labels in gitbot}.
In: \bbtitle{Proceedings of the Fourth International Workshop on Bots in Software Engineering},
pp. \bfpage{28}--\blpage{32}
(\byear{2022})
\end{bchapter}
\endbibitem

%%% 53
\bibitem[\protect\citeauthoryear{Zolkeply and Shao}{2019}]{zolkeply2019classifying}
\begin{bchapter}
\bauthor{\bsnm{Zolkeply}, \binits{M.S.}},
\bauthor{\bsnm{Shao}, \binits{J.}}:
\bctitle{Classifying software issue reports through association mining}.
In: \bbtitle{Proceedings of the 34th ACM/SIGAPP Symposium on Applied Computing},
pp. \bfpage{1860}--\blpage{1863}
(\byear{2019})
\end{bchapter}
\endbibitem

%%% 54
\bibitem[\protect\citeauthoryear{Zhou et~al.}{2016}]{zhou2016combining}
\begin{barticle}
\bauthor{\bsnm{Zhou}, \binits{Y.}},
\bauthor{\bsnm{Tong}, \binits{Y.}},
\bauthor{\bsnm{Gu}, \binits{R.}},
\bauthor{\bsnm{Gall}, \binits{H.}}:
\batitle{Combining text mining and data mining for bug report classification}.
\bjtitle{Journal of Software: Evolution and Process}
\bvolume{28}(\bissue{3}),
\bfpage{150}--\blpage{176}
(\byear{2016})
\end{barticle}
\endbibitem

%%% 55
\bibitem[\protect\citeauthoryear{Devine et~al.}{2023}]{devine2023evaluating}
\begin{barticle}
\bauthor{\bsnm{Devine}, \binits{P.}},
\bauthor{\bsnm{Koh}, \binits{Y.S.}},
\bauthor{\bsnm{Blincoe}, \binits{K.}}:
\batitle{Evaluating software user feedback classifier performance on unseen apps, datasets, and metadata}.
\bjtitle{Empirical Software Engineering}
\bvolume{28}(\bissue{2}),
\bfpage{26}
(\byear{2023})
\end{barticle}
\endbibitem

%%% 56
\bibitem[\protect\citeauthoryear{Ebrahim and Joy}{2024}]{ebrahim2024few}
\begin{bchapter}
\bauthor{\bsnm{Ebrahim}, \binits{F.}},
\bauthor{\bsnm{Joy}, \binits{M.}}:
\bctitle{Few-shot issue report classification with adapters}.
In: \bbtitle{Proceedings of the Third ACM/IEEE International Workshop on NL-based Software Engineering},
pp. \bfpage{41}--\blpage{44}
(\byear{2024})
\end{bchapter}
\endbibitem

%%% 57
\bibitem[\protect\citeauthoryear{Kumar et~al.}{2023}]{kumar2023github}
\begin{bchapter}
\bauthor{\bsnm{Kumar}, \binits{G.S.}},
\bauthor{\bsnm{Angel}, \binits{T.S.}},
\bauthor{\bsnm{Chakreborthy}, \binits{S.}},
\bauthor{\bsnm{Reddy}, \binits{K.D.}}:
\bctitle{Github bug classification using pipeline approach in machine learning}.
In: \bbtitle{2023 International Conference on Data Science, Agents \& Artificial Intelligence (ICDSAAI)},
pp. \bfpage{1}--\blpage{7}
(\byear{2023}).
\bcomment{IEEE}
\end{bchapter}
\endbibitem

%%% 58
\bibitem[\protect\citeauthoryear{Bharadwaj and Kadam}{2022}]{bharadwaj2022github}
\begin{bchapter}
\bauthor{\bsnm{Bharadwaj}, \binits{S.}},
\bauthor{\bsnm{Kadam}, \binits{T.}}:
\bctitle{Github issue classification using bert-style models}.
In: \bbtitle{Proceedings of the 1st International Workshop on Natural Language-based Software Engineering},
pp. \bfpage{40}--\blpage{43}
(\byear{2022})
\end{bchapter}
\endbibitem

%%% 59
\bibitem[\protect\citeauthoryear{Kwak and Lee}{2022}]{kwak2022issue}
\begin{botherref}
\oauthor{\bsnm{Kwak}, \binits{C.}},
\oauthor{\bsnm{Lee}, \binits{S.}}:
Issue report classification using a multimodal deep learning technique
(2022)
\end{botherref}
\endbibitem

%%% 60
\bibitem[\protect\citeauthoryear{Du et~al.}{2024}]{du2024llm}
\begin{botherref}
\oauthor{\bsnm{Du}, \binits{X.}},
\oauthor{\bsnm{Liu}, \binits{Z.}},
\oauthor{\bsnm{Li}, \binits{C.}},
\oauthor{\bsnm{Ma}, \binits{X.}},
\oauthor{\bsnm{Li}, \binits{Y.}},
\oauthor{\bsnm{Wang}, \binits{X.}}:
Llm-brc: A large language model-based bug report classification framework.
Software Quality Journal,
1--21
(2024)
\end{botherref}
\endbibitem

%%% 61
\bibitem[\protect\citeauthoryear{Shen et~al.}{2024}]{shen2024mitigating}
\begin{barticle}
\bauthor{\bsnm{Shen}, \binits{J.}},
\bauthor{\bsnm{Li}, \binits{Z.}},
\bauthor{\bsnm{Lu}, \binits{Y.}},
\bauthor{\bsnm{Pan}, \binits{M.}},
\bauthor{\bsnm{Li}, \binits{X.}}:
\batitle{Mitigating the impact of mislabeled data on deep predictive models: an empirical study of learning with noise approaches in software engineering tasks}.
\bjtitle{Automated Software Engineering}
\bvolume{31}(\bissue{1}),
\bfpage{33}
(\byear{2024})
\end{barticle}
\endbibitem

%%% 62
\bibitem[\protect\citeauthoryear{GayathriP and Babu}{2019}]{GayathriP2019OptimizationOT}
\begin{bchapter}
\bauthor{\bsnm{GayathriP}, \binits{M.}},
\bauthor{\bsnm{Babu}, \binits{G.K.K.}}:
\bctitle{Optimization of the bug report classification using genetic algorithm}.
(\byear{2019}).
\burl{https://api.semanticscholar.org/CorpusID:212538218}
\end{bchapter}
\endbibitem

%%% 63
\bibitem[\protect\citeauthoryear{Li et~al.}{2022}]{li2022robust}
\begin{bchapter}
\bauthor{\bsnm{Li}, \binits{Z.}},
\bauthor{\bsnm{Pan}, \binits{M.}},
\bauthor{\bsnm{Pei}, \binits{Y.}},
\bauthor{\bsnm{Zhang}, \binits{T.}},
\bauthor{\bsnm{Wang}, \binits{L.}},
\bauthor{\bsnm{Li}, \binits{X.}}:
\bctitle{Robust learning of deep predictive models from noisy and imbalanced software engineering datasets}.
In: \bbtitle{Proceedings of the 37th IEEE/ACM International Conference on Automated Software Engineering},
pp. \bfpage{1}--\blpage{13}
(\byear{2022})
\end{bchapter}
\endbibitem

%%% 64
\bibitem[\protect\citeauthoryear{Rejithkumar et~al.}{2024}]{rejithkumar2024text}
\begin{bchapter}
\bauthor{\bsnm{Rejithkumar}, \binits{G.}},
\bauthor{\bsnm{Anish}, \binits{P.R.}},
\bauthor{\bsnm{Ghaisas}, \binits{S.}}:
\bctitle{Text-to-text generation for issue report classification}.
In: \bbtitle{Proceedings of the Third ACM/IEEE International Workshop on NL-based Software Engineering},
pp. \bfpage{53}--\blpage{56}
(\byear{2024})
\end{bchapter}
\endbibitem

%%% 65
\bibitem[\protect\citeauthoryear{Du et~al.}{2017}]{du2017automatic}
\begin{bchapter}
\bauthor{\bsnm{Du}, \binits{X.}},
\bauthor{\bsnm{Zheng}, \binits{Z.}},
\bauthor{\bsnm{Xiao}, \binits{G.}},
\bauthor{\bsnm{Yin}, \binits{B.}}:
\bctitle{The automatic classification of fault trigger based bug report}.
In: \bbtitle{2017 IEEE International Symposium on Software Reliability Engineering Workshops (ISSREW)},
pp. \bfpage{259}--\blpage{265}
(\byear{2017}).
\bcomment{IEEE}
\end{bchapter}
\endbibitem

%%% 66
\bibitem[\protect\citeauthoryear{Fan et~al.}{2017}]{fan2017road}
\begin{bchapter}
\bauthor{\bsnm{Fan}, \binits{Q.}},
\bauthor{\bsnm{Yu}, \binits{Y.}},
\bauthor{\bsnm{Yin}, \binits{G.}},
\bauthor{\bsnm{Wang}, \binits{T.}},
\bauthor{\bsnm{Wang}, \binits{H.}}:
\bctitle{Where is the road for issue reports classification based on text mining?}
In: \bbtitle{2017 ACM/IEEE International Symposium on Empirical Software Engineering and Measurement (ESEM)},
pp. \bfpage{121}--\blpage{130}
(\byear{2017}).
\bcomment{IEEE}
\end{bchapter}
\endbibitem

%%% 67
\bibitem[\protect\citeauthoryear{Meng and Visser}{2023}]{meng2023bug}
\begin{bchapter}
\bauthor{\bsnm{Meng}, \binits{Q.}},
\bauthor{\bsnm{Visser}, \binits{J.}}:
\bctitle{Which bug reports are valid and why? using the bert transformer to classify bug reports and explain their validity}.
In: \bbtitle{Proceedings of the 4th European Symposium on Software Engineering},
pp. \bfpage{52}--\blpage{60}
(\byear{2023})
\end{bchapter}
\endbibitem

%%% 68
\bibitem[\protect\citeauthoryear{Rahman et~al.}{2022}]{rahman2022works}
\begin{barticle}
\bauthor{\bsnm{Rahman}, \binits{M.M.}},
\bauthor{\bsnm{Khomh}, \binits{F.}},
\bauthor{\bsnm{Castelluccio}, \binits{M.}}:
\batitle{Works for me! cannot reproduce--a large scale empirical study of non-reproducible bugs}.
\bjtitle{Empirical Software Engineering}
\bvolume{27}(\bissue{5}),
\bfpage{111}
(\byear{2022})
\end{barticle}
\endbibitem

%%% 69
\bibitem[\protect\citeauthoryear{Trautsch and Herbold}{2022}]{trautsch2022predicting}
\begin{bchapter}
\bauthor{\bsnm{Trautsch}, \binits{A.}},
\bauthor{\bsnm{Herbold}, \binits{S.}}:
\bctitle{Predicting issue types with sebert}.
In: \bbtitle{Proceedings of the 1st International Workshop on Natural Language-based Software Engineering},
pp. \bfpage{37}--\blpage{39}
(\byear{2022})
\end{bchapter}
\endbibitem

%%% 70
\bibitem[\protect\citeauthoryear{Izadi}{2022}]{izadi2022catiss}
\begin{bchapter}
\bauthor{\bsnm{Izadi}, \binits{M.}}:
\bctitle{Catiss: An intelligent tool for categorizing issues reports using transformers}.
In: \bbtitle{Proceedings of the 1st International Workshop on Natural Language-based Software Engineering},
pp. \bfpage{44}--\blpage{47}
(\byear{2022})
\end{bchapter}
\endbibitem

\end{thebibliography}

\newpage
\appendix
\section{A map of automatic techniques for issue report classification}\label{appendix:complete-map}
\begin{scriptsize}
\begin{longtable}{p{0.6cm}p{2.2cm}p{1cm}p{2.5cm}p{1.8cm}p{1.5cm}p{1.6cm}}
    \caption{A map of automatic techniques for issue report classification. Precision (P), Recall (R), F-score (F1), Accuracy (Acc), Micro Average (Avg), Test on a fixed set (TFS), and 10-fold CV (10FCV)}
\label{tab:map-of-SA-to-SE}
\\ \toprule
  \textbf{PS} & \textbf{Technique} & \textbf{Data source} & \textbf{Features} & \textbf{Pre-processing technique} & \textbf{Evaluation strategy} & \textbf{Evaluation metrics}
 \\ \toprule
\endfirsthead
%\toprule 
%
\multicolumn{7}{l}{\footnotesize{\textit{\tablename\ \thetable\ continued from previous page}}} \\
\midrule % Used hline instead of midrule for less space to the "...continued from..." line; see above.
  \textbf{PS} & \textbf{Technique} & \textbf{Data source} & \textbf{Features} & \textbf{Pre-processing technique} & \textbf{Evaluation strategy} & \textbf{Evaluation metrics}
\\
%\toprule
\midrule
\endhead
\hline % Used hline instead of midrule for less space to the "Continued on..." line; see below.
\multicolumn{7}{r}{\footnotesize{\textit{Continued on next page}}} 
\\
%\toprule
\endfoot
%\toprule
%\bottomrule
\endlastfoot
        PS1 & Rule-based technique & Bugzilla & Title, description & TF-IDF & TFS & P, R \\ \midrule
        PS2 & Random forest, SVM, K-means clustering  & Bugzilla, Jira & Summary & TF, TF-IDF, TF-IGM & TFS & R, P, F1 \\ \midrule
        PS3 & BERT-based model, RoBERTa, CodeBERT, BERTOverflow,  seBERT & GitHub & Title, description & BERT-based tokenizer & TFS & Micro F1 \\ \midrule
        PS4 & CNN & GitHub & Image, code, & Embedding layer & TFS & P, R, F1 \\ \midrule
        PS5 & KNN & Bugzilla, Redhat Bugzilla & Summary, severity, priority, component, assignee, reporter, OS, reproducibility & TF-IDF & TFS & R, P, F1 \\ \midrule
        PS6 & Stacking ensemble, Random forest, SVM, Logistic regression, Multilayer perceptron, SGD, Multinomial Naive Bayes, Naive Bayes & GitHub, Bugzilla, Jira & Title, description & TF-IDF & Stratified 10FCV & Acc, MCC, F1 \\ \midrule
        PS7 & SGD & GitHub & Title, body & TF-IDF & TFS & P, R,  F1,  Avg P, Avg R , Avg F1 \\ \midrule
        PS8 & GPT-3.5-turbo base model & GitHub & Title, body & GPT-based tokenizer & TFS & P, R,  F1,  Avg P, Avg R , Avg F1 \\ \midrule
        PS9 & Random tree, SVM, Naive Bayes & GitHub, Bugzilla & Title, description & Manual selection of textual features & 5-fold CV & Acc \\ \midrule
        PS10 & Naive Bayes, SVM, KNN, Logistic regression, Decision tree, Random forest & CSS data & Description & TF-IDF & N-fold CV & F1 \\ \midrule
        PS11 & RoBERTa & GitHub & Title, body & RoBERTa-based tokenizer & TFS & P, R, F1 \\ \midrule
        PS12 & BERT-based model & GitHub & Title, body & BERT-based tokenizer & TFS & P, R,  F1,  Avg P, Avg R , Avg F1 \\ \midrule
        PS13 & Logistic regression, Random forest & Jira & Title, description & N-gram IDF & 10FCV, TFS & F1 \\ \midrule
        PS14 & Vector space model-based approach & NA & Description & TF-IDF & NA & Acc, P, R \\ \midrule
        PS15 & SVM & Bugzilla & Features based on collaboration graphs & NA & TFS & P, R, F1 \\ \midrule
        PS16 & SVM, Random forest & Bugzilla & Title, description, comments, reporter experience, collaboration network, completeness, readability & TF-IDF & Time split with 10FCV & P, R, F1,  AUC \\ \midrule
        PS17 & RoBERTa & GitHub & Title, body & RoBERTa-based tokenizer & TFS & P, R,  F1,  Avg P, Avg R , Avg F1 \\ \midrule
        PS18 & BERT-based model & GitHub & Title, description, submitter & BERT-based tokenizer & TFS & P, R, F1 \\ \midrule
        PS19 & LSTM & Bugzilla, Jira & Title, description & Skip-gram model & Time split with test on a fixed set  & P, R, F1 \\ \midrule
        PS20 & Decision tree, Naive Bayes, and Logistic regression. & Jira & Title, description, discussion & LDA & 10FCV & F1 \\ \midrule
        PS21 & CNN, LSTM & GitHub & Title, body, user manuals & Embedding layer, GloVe, Skip-gram model & 5-fold CV & P, R, F1 \\ \midrule
        PS22 & Fasttext & GitHub & Title, body & Skip-gram model & 10FCV & P, R, F1 \\ \midrule
        PS23 & Classification associations rule mining & Bugzilla, Jira & Title, description & Manuall selection of textual features & TFS & Acc \\ \midrule
        PS24 & Naive Bayes, ADTree, Logistic regression & Mantis, Redhat Bugzilla & Summary, severity, priority, component, assignee, reporter, OS, reproducibility & NA & 10FCV & P, R, F1 \\ \midrule
        PS25 & CNN & Bugzilla & Summary, description & Skip-gram model & Time split with test on a fixed set  & P, R, F1,  AUC \\ \midrule
        PS26 & KNN, SVM, Logistic regression, Decision tree & CSS data & Heading, description,  priority, submitter experience & TF-IDF & Time split with 10FCV & AUC, MCC, P, R, F1 \\ \midrule
        PS27 & Distilbert, art-large-mnli & Twitter, Other forum posts & Review description, Metadata-based features, e.g., app rating and category & BERT-based tokenizer & 5-fold CV & P, R, F1 \\ \midrule
        PS28 & RoBERTa with adapters & GitHub & Title, body & RoBERTa-based tokenizer & TFS & P, R,  F1,  Avg P, Avg R , Avg F1 \\ \midrule
        PS29 & Setfit & GitHub & Title, body & All-mpnet-base-v2 & TFS & P, R,  F1,  Avg P, Avg R , Avg F1 \\ \midrule
        PS30 & Naive Bayes, Logistic regression, Decision tree & GitHub & Description & TF-IDF & TFS & Acc, P,  R \\ \midrule
        PS31 & BERT-based model & GitHub & Title, body & BERT-based tokenizer & TFS & P, R,  F1,  Avg P, Avg R , Avg F1 \\ \midrule
        PS32 & BERT-based model, XGBoost, SVM, Logistic regression, SVM, DistillBert, CNN, Random forest  & CSS data & Heading, description, priority, submitter experience, completeness & Skip-gram model & Time split with 10FCV & P, R, F1, Acc, AUC, MCC \\ \midrule
        PS33 & CNN & GitHub & Image, code, & Embedding layer & TFS & P, R, F1 \\ \midrule
        PS34 & BERT-based model, RoBERTa, ALBERT & GitHub & Title, body, issue-author association & AutoTokenizer from transformers & TFS & P, R,  F1,  Avg P, Avg R , Avg F1 \\ \midrule
        PS35 & Setfit, GPT3.5-turbo & GitHub & Title, body & All-mpnet-base-v2 & TFS & P, R, F1, F1-micro \\ \midrule
        PS36 & Feed-forward neural network & GitHub & Title, description, comment & Text-embedding-ada-002 & 5-fold CV & Acc, P, R, F1 \\ \midrule
        PS37 & LSTM, CNN & Bugzilla, Jira & Title, description & NA & 10FCV & F-score, AUC, MCC, G-measure \\ \midrule
        PS38 & KNN & Bugzilla & Summary, severity, component, priority & TF-IDF & TFS & Acc \\ \midrule
        PS39 & LSTM & Bugzilla, Jira & Title, body & Skip-gram model & 10FCV & AUC, MCC, G-measure, F1 \\ \midrule
        PS40 & T5 & GitHub & Title, body & T5-based tokenizer & TFS & P, R,  F1,  Avg P, Avg R , Avg F1 \\ \midrule
        PS41 & SGD, Logistic regression, Naive Bayes, Gradient boosting, Decision tree, Random forest, Linear discriminant analysis & Bugzilla & Summary, description & Skip-gram model & TFS & Acc, F1 \\ \midrule
        PS42 & SVM, Random forest, Logistic regression, Naive Bayes & GitHub & Title, body & TF-IDF & 10FCV & P, R, F1 \\ \midrule
        PS43 & BERT-based model & Bugzilla & Title, description & BERT-based tokenizer & TFS & P, R, F1 \\ \midrule
        PS44 & XGBoost, Naive Bayes, Logistic regression, Random forest, J48 & Bugzilla & Priority, severity, component, text readability, text sentiment & NA & 10FCV & P, R, F1 \\ \midrule
        PS45 & seBERT & GitHub & Title, body & BERT-based tokenizer & TFS & P, R,  F1,  Avg P, Avg R , Avg F1 \\ \midrule
        PS46 & RoBERTa & GitHub & Title, body & RoBERTa-based tokenizer & TFS & P, R,  F1,  Avg P, Avg R , Avg F1 \\ \bottomrule

\end{longtable}
\end{scriptsize}
\section{List of included primary studies}

\begin{scriptsize}
\begin{longtable}{p{1.7cm}p{10cm}p{1.5cm}}
    \caption{List of included primary studies}
    \label{tab:included-PSs}
\\ \toprule
\textbf{Id} & \textbf{Title} & \textbf{Ref.}
 \\ \toprule
\endfirsthead
\toprule 

\textbf{Id \newline(continued)} & \textbf{Title (continued)} & \textbf{Ref.\newline (continued)}
\\
\toprule
\endhead
\toprule
\endfoot
\toprule
\endlastfoot

PS1 & A bug rule based technique with feedback for classifying bug reports & \cite{zhang2011bug}
\\ \midrule

PS2 & A Comparative Study of Short Text Classification Methods for Bug Report Type Identification
 & \cite{polpinij2022comparative}
\\ \midrule

PS3 & A Comparison of Pretrained Models for Classifying Issue Reports & \cite{heo2024comparison}
\\ \midrule

PS4 & A Multimodal Deep Learning Model Using Text, Image, and Code Data for Improving Issue Classification Tasks
 & \cite{kwak2023multimodal}
\\ \midrule

PS5 & A Supervised Bug Report Classification with Incorporate and Textual field Knowledge
 & \cite{kukkar2018supervised}
\\ \midrule

PS6 & An empirical evaluation of stacked generalization models for binary bug report classification
 & \cite{alraddadi2024empirical}
\\ \midrule

PS7 & An Intelligent Tool for Classifying Issue Reports
 & \cite{laiq2023intelligent}
\\ \midrule

PS8 & Applying Large Language Models to Issue Classification
 & \cite{aracena2024applying}
\\ \midrule

PS9 & Automated classification of software bug reports
 & \cite{otoom2019automated}
\\ \midrule

PS10 & Automated classification of unstructured bilingual software bug reports: An industrial case study research
 & \cite{koksal2021automated}
\\ \midrule

PS11 & Automatic Issue Classifier: A Transfer Learning Framework for Classifying Issue Reports
 & \cite{nadeem2021automatic}
\\ \midrule

PS12 & BERT-Based GitHub Issue Report Classification
 & \cite{siddiq2022bert}
\\ \midrule

PS13 & Bug or not? Bug Report classification using N-gram IDF
 & \cite{terdchanakul2017bug}
\\ \midrule

PS14 & Bug report classification based on vector space model
& \cite{chen2019bug}
\\ \midrule

PS15 & Categorizing bugs with social networks: A case study on four open source software communities
 & \cite{zanetti2013categorizing}
\\ \midrule

PS16 & Chaff from the Wheat: Characterizing and Determining Valid Bug Reports
 & \cite{fan2018chaff}
\\ \midrule

PS17 & ClassifAI: Automating Issue Reports Classification using Pre-Trained BERT (Bidirectional Encoder Representations from Transformers) Language Models
 & \cite{alam2024classifai}
\\ \midrule

PS18 & Classification of open source software bug report based on transfer learning
 & \cite{zhifang2024classification}
\\ \midrule

PS19 & Classifying bug reports into bugs and non-bugs using LSTM
 & \cite{qin2018classifying}
\\ \midrule

PS20 & Classifying bug reports to bugs and other requests using topic modeling
& \cite{pingclasai2013classifying}
\\ \midrule

PS21 & Classifying issue reports according to feature descriptions in a user manual based on a deep learning model
& \cite{cho2022classifying}
\\ \midrule

PS22 & Classifying Issues into Custom Labels in GitBot
 & \cite{park2022classifying}
\\ \midrule

PS23 & Classifying software issue reports through association mining
 & \cite{zolkeply2019classifying}
\\ \midrule

PS24 & Combining text mining and data mining for bug report classification
 & \cite{zhou2016combining}
\\ \midrule

PS25 & Deep learning based valid bug reports determination and explanation
 & \cite{he2020deep}
\\ \midrule

PS26 & Early Identification of Invalid Bug Reports in Industrial Settings – A Case Study
 & \cite{laiq2022early}
\\ \midrule

PS27 & Evaluating software user feedback classifier performance on unseen apps, datasets, and metadata
 & \cite{devine2023evaluating}
\\ \midrule

PS28 & Few-Shot Issue Report Classification with Adapters & \cite{ebrahim2024few}
\\ \midrule

PS29 & Few-Shot Learning for Issue Report Classification & \cite{colavito2023few}
\\ \midrule

PS30 & GitHub Bug Classification Using Pipeline Approach in Machine Learning & \cite{kumar2023github}
\\ \midrule

PS31 & GitHub Issue Classification Using BERT-Style Models & \cite{bharadwaj2022github}
\\ \midrule

PS32 & Industrial adoption of machine learning techniques for early identification of invalid bug reports & \cite{laiq2024industrial}
\\ \midrule

PS33 & Issue report classification using a multimodal deep learning technique & \cite{kwak2022issue}
\\ \midrule

PS34 & Issue Report Classification Using Pre-trained Language Models & \cite{colavito2022issue}
\\ \midrule

PS35 & Large Language Models for Issue Report Classification & \cite{colavito2024large}
\\ \midrule

PS36 & LLM-BRC: A large language model-based bug report classification framework & \cite{du2024llm}
\\ \midrule

PS37 & Mitigating the impact of mislabeled data on deep predictive models: an empirical study of learning with noise approaches in software engineering tasks & \cite{shen2024mitigating}
\\ \midrule

PS38 & Optimization of the bug report classification using genetic algorithm &  \cite{GayathriP2019OptimizationOT}
\\ \midrule

PS39 & Robust Learning of Deep Predictive Models from Noisy and Imbalanced Software Engineering Datasets & \cite{li2022robust}
\\ \midrule

PS40 & Text-To-Text Generation for Issue Report Classification & \cite{rejithkumar2024text}
\\ \midrule

PS41 & The automatic classification of fault trigger based bug report & \cite{du2017automatic}
\\ \midrule

PS42 & Where Is the Road for Issue Reports Classification Based on Text Mining? & \cite{fan2017road}
\\ \midrule

PS43 & Which bug reports are valid and why? Using the BERT transformer to classify bug reports and explain their validity & \cite{meng2023bug}
\\ \midrule

PS44 & Works for Me! Cannot Reproduce – A Large Scale Empirical Study of Non-reproducible Bugs & \cite{rahman2022works}
\\ \midrule

PS45 & Predicting Issue Types with seBERT & \cite{trautsch2022predicting}
\\ \midrule

PS46 & CatIss: An Intelligent Tool for Categorizing Issues reports using Transformers & \cite{izadi2022catiss}
\\ \bottomrule

\end{longtable}
\end{scriptsize}

\end{document}